\documentclass[12pt, draftclsnofoot, onecolumn]{IEEEtran}
\usepackage{graphicx}
\usepackage{caption}
\ifCLASSOPTIONcompsoc
  \usepackage[caption=false,font=normalsize,labelfont=sf,textfont=sf]{subfig}
\else
  \usepackage[caption=false,font=footnotesize]{subfig}
\fi
\usepackage{indentfirst}

\usepackage{amsmath}
\allowdisplaybreaks[3]
\usepackage{amssymb}
\usepackage{times}
\usepackage{mathtools}
\usepackage{psfrag}
\usepackage{cite}
\usepackage{lastpage}
\usepackage{fancyhdr}
\usepackage{color}
 \usepackage{amsthm}
\usepackage{bigints}
\newtheorem{theorem}{Theorem}

\newtheorem{Assumption}{Assumption}
\newtheorem{Lemma}{Lemma}

\newtheorem{Corollary}{Corollary}
\begin{document}
\title{On the Outage Performance of Network NOMA (N-NOMA)  Modeled {\color{black}by} Poisson Line Cox Point Process}
\author{Yanshi Sun, \IEEEmembership{Student Member, IEEE}, Zhiguo Ding, \IEEEmembership{Senior Member, IEEE}, Xuchu Dai,
\thanks{Y. Sun and X. Dai are with the Key Laboratory of Wireless-Optical Communications, Chinese Academy of Sciences, School of Information Science and Technology, University of Science and Technology of China,
No. 96 Jinzhai Road, Hefei, Anhui Province, 230026, P. R. China. (email: sys@mail.ustc.edu.cn, daixc@ustc.edu.cn).

Z. Ding is with the School of Electrical and Electronic Engineering, the University of Manchester, Manchester M13 9PL, U.K. (email:zhiguo.ding@manchester.ac.uk).
}}
\maketitle

\begin{abstract}
This paper studies the application of network non-orthogonal multiple access (N-NOMA) to vehicular communication networks, modeled by {\color{black}a} Poisson line Cox point process (PLC). Particularly, the road infrastructure is modeled by a Poisson line point process (PLP), {\color{black}whereas} base stations (BSs) and users are {\color{black}located} according to 1D homogeneous Poisson point processes (HPPPs) on each road. In the considered N-NOMA scheme, {\color{black}there are two types of users, i.e., coordinated multi-point (CoMP) users which are far from their BSs and NOMA users which are close to their BSs, where the CoMP users and the NOMA users are served simultaneously by applying the principle of NOMA}. Specifically, {\color{black}the} BSs {\color{black}collaborate with each other} to provide reliable transmission for  the CoMP user {\color{black}by applying Almouti coding}, while each BS employs superposition coding to opportunistically serve an additional NOMA user. Compared to conventional OMA based scheme, where only the CoMP user is served, the proposed N-NOMA scheme significantly improves the spectral efficiency and {\color{black}user connectivity}, since more users are served in the {\color{black}same} resource block. A comprehensive analytical framework is developed to characterize {\color{black}the} system level performance of the proposed N-NOMA scheme {\color{black}by applying PLC}. Closed form expressions for the outage probabilities achieved by the CoMP and NOMA users are obtained. Computer simulation results are provided to show the superior performance of the proposed scheme and also validate the accuracy of the developed analytical results.
\end{abstract}
\section{Introduction}
Recently, non-orthogonal multiple access (NOMA) has been recognized as a key enabling technique for future communication networks, due to its superior spectral efficiency and {\color{black}capability to support} massive connectivity \cite{saito2013non,ding2017NOMAsurvey,islam2017noma5G, ding2017application}. The key idea of NOMA is to encourage multiple users to occupy the same resource block simultaneously and apply {\color{black}advanced multi-user detection} techniques such as successive interference cancellation (SIC) to mitigate inter-user interferences. It is shown in the literature that NOMA is compatible with many other advanced communication techniques, such as massive multiple-input multiple-output (MIMO) techniques \cite{ding2016general,choi2015minimum,octavia2017MIMONOMA}, millimeter-wave networks
\cite{ding2017randommmwave,yzhoummwvave2018,sysmmwave2018}, and mobile edge computing (MEC) \cite{Ding2019MEC}, {\color{black}etc}.

 {\color{black}As an important form of NOMA, network} NOMA (N-NOMA) \cite{choi2014non,ali2018coordinated,tian2016performance} has been recently proposed by applying NOMA to {\color{black}conventional} network MIMO systems \cite{venkatesan2007network,huang2009increasing}, {\color{black}i.e.}, coordinated multi-point (CoMP) systems \cite{jungnickel2009coordinated}, to enhance the performance of network MIMO. Note that in traditional network MIMO systems,
users are served by multiple base stations (BSs) through cooperative transmission, {\color{black}and such users are denoted by CoMP users in this paper}. In N-NOMA, {\color{black}in addition to those CoMP users}, each BS individually  serves additional near users by using the same resource blocks {\color{black}which would have been solely occupied by the CoMP users}.
In \cite{choi2014non} and \cite{sys2017nnomafeasibility}, Alamouti coding and analog beamforming {\color{black}have been} applied to {\color{black}improve} the cell-edge user's reception reliability, respectively. In \cite{ali2018downlink}, power allocation for downlink N-NOMA was studied. In \cite{sys2019PCP}, the application of N-NOMA to uplink CoMP systems {\color{black}has been} investigated.

This paper studies the application of N-NOMA to  vehicular communication networks. To meet the rapid development of intelligent transport {\color{black}systems} (ITSs), the concept of vehicle-to-everything (V2X) has been proposed recently \cite{sahin2018virtual,boban2018connected,chen2017vehicle}. According to the standardization work by the 3rd Generation Partnership Project
(3GPP) \cite{GPP2016V2X}, V2X communication {\color{black}includes} vehicle-to-vehicle (V2V), vehicle-to-infrastructure (V2I),  vehicle-to-pedestrian (V2P) communications, etc.  Due to the ever-increasing user density and the higher quality of service (QoS) requested by per user in vehicular networks, the requirements of reliable data transmission and low latency cannot be satisfied by merely applying  traditional OMA based multiple access techniques. Hence, NOMA based schemes are considered in V2X networks to support more vehicles by occupying the limited spectrum resources.
For example, the application of NOMA {\color{black}assisted by} spatial modulation (SM) to V2V networks was investigated in \cite{chen2017performance}.
The scheduling and time frequency  resource allocation problem in a V2X broadcasting system  was studied in \cite{di2017non}. The performance of full duplex NOMA (FD-NOMA) in a decentralized V2X system was characterized in \cite{zhang2019performance}.
However, the previous studies on NOMA in V2X networks {\color{black}mainly} focused on {\color{black}the case in which} a user can only belong to a single NOMA group and cooperation between multiple BSs to serve a particular user was not considered. This is not {\color{black}applicable to} scenarios {\color{black}with}  users very far from BSs. Thus, the study of applying N-NOMA to V2X networks is necessary and urgent, which is the first  motivation of this paper.  Besides, the previous studies on NOMA in V2X networks {\color{black}do not} take the topology of the whole vehicular network, {\color{black}i.e., vehicles are constrained by road infrastructure}, into consideration, and hence {\color{black}there is a lack of} rigorous characterization of the interference. To fulfill this blank, another motivation of this paper is to provide {\color{black}the} system level performance of N-NOMA  by taking the topology of the vehicular network into consideration. The contributions of this paper are listed as follows:

\begin{itemize}
	\item To capture the topology of the vehicular network, a doubly stochastic spatial model called Poisson line Cox point process (PLC) \cite{choi2018poisson,choi2018analytical,chetlur2018coverage} is considered. Particularly, the layout of roads is modeled according to a Poisson line point process (PLP). BSs and users are modeled as 1D homogeneous Poisson point processes (HPPPs)  on each road. Two kinds of users are considered, namely the CoMP users which are far from {\color{black}the} BSs and the NOMA users which are close to {\color{black}the} BSs. N-NOMA is applied to serve the CoMP users and NOMA users by using the same resource block simultaneously. Alamouti coding is applied to provide diversity gain for the CoMP users, while superposition coding is applied at each BS to include additional NOMA users.
	\item Outage probability is used as a metric to characterize the performance of the proposed N-NOMA scheme. One key step of the analysis is to characterize the interference from other wireless nodes in the vehicular network. Since Alamouti coding is applied in the proposed N-NOMA scheme, we need to utilize  two {\color{black}consecutive} time slots to transmit the intended signals. A problem arising here is that the interferences observed at the two time slots are correlated, due to the fact that N-NOMA is also applied by some interfering nodes in the whole network. Hence, the situation in this paper is more complicated than that in the conventional stochastic {\color{black}geometric}
	 analysis, where only one time slot is considered. However, it is proved that the aggregated interference which is obtained by taking weighted sums over the two time slots, is equivalent to the interference observed in one time slot. By using this interesting observation, the Laplace transform of the interference is derived, based on which closed form expressions for the outage probabilities achieved by the CoMP and NOMA users are obtained.
	 \item Asymptotic analysis for the case when the density of roads {\color{black}goes to infinity} and the density of BSs on each road {\color{black}goes to zero} is also provided. In this case, it is found that the distribution of the interfering nodes in the considered PLC approaches a {\color{black}conventional} 2D HPPP.
	\item The developed analytical results are verified by computer simulations. Comparison between the proposed N-NOMA and conventional OMA {\color{black}for the case with one CoMP user} is provided. To get more insight into the proposed scheme, the impact of the system parameters, such as density of roads and nodes on each road, power allocation coefficients, distances between the BSs and users, on the performance of the proposed N-NOMA scheme, is demonstrated and discussed.
\end{itemize}

The rest of the paper is organized as follows. Section II provides the illustration of spatial modeling of the considered vehicular communication system and the description of the proposed N-NOMA scheme. Section III discusses the decoding of the signals for both CoMP and NOMA users.  Section IV analyzes the performance of the proposed N-NOMA scheme. Section V provides numerical results to verify the accuracy of the developed analytical results and to demonstrate the performance of the proposed scheme. Section VI concludes the paper. Finally, appendices collect the proofs of the obtained analytical results.
\section{System Model}
\begin{figure}[!t]
\centering
\includegraphics[width=3in]{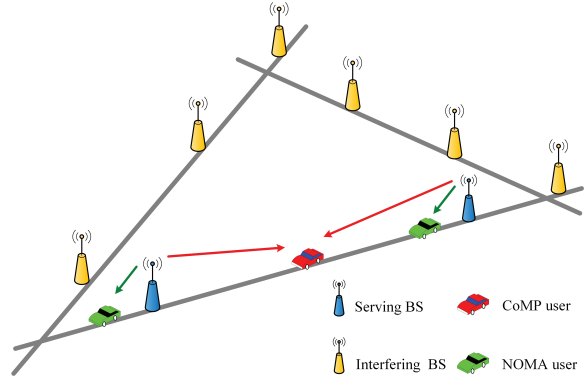}
\caption{An illustration of the system model.}
\label{system_model}
\end{figure}
\subsection{Poisson line point process (PLP)}
A PLP $\Phi_l$ with parameter $\lambda_l$ is determined by a HPPP $\Xi$ with intensity $\lambda_l$ on the representation space
$\mathcal{C}:=\mathbb{R}^+\times[0,2\pi)$. Specifically, any line $l_i$ in $\Phi_l$ can be parameterized by a unique point of $\Xi$, denoted by $(\rho_i,\theta_i)$, according to the following
equation
\begin{align}
	l_i=\{(x,y)\in\mathbb{R}^2|x\cos{(\theta_i)}+y\sin{(\theta_i)}=\rho_i\},
\end{align}
where $\rho_i$ is the distance from the origin to line $l_i$, and $\theta_i$ is the angle between the perpendicular onto line $l_i$  and the positive x-axis measured in a counterclockwise direction.
\subsection{Poisson line Cox point process (PLC)}
Given a PLP $\Phi_l$ with parameter $\lambda_l$, a PLC $\bar{\Phi}_t(\lambda_l,\lambda_t)$ with parameter $\lambda_l$ and $\lambda_t$ can be produced by dropping points
onto all lines of $\Phi_l$, according to a 1D HPPP with the same intensity $\lambda_t$ on each line. Mathematically,
\begin{align}
 \bar{\Phi}_t(\lambda_l,\lambda_t):=\{\Psi_{l_i}\}_{l_i\in\Phi_l},
\end{align}
where $\Psi_{l_i}$ is a 1D HPPP with intensity $\lambda_t$ on line $l_i$.
\subsection{Spatial modeling of the BSs and users}
This paper focuses on a vehicular network {\color{black}consisting of} BSs and users {\color{black}(vehicles)} which are restricted to roadways. To begin with, the road system is modeled by a PLP $\Phi_l$ driven by a HPPP $\Xi$ with intensity $\lambda_l$. Denote the $i$-th road by $l_i$, then $\Phi_l=\{l_i\}$ is the collection of all roads. Given the PLP $\Phi_l$, the BSs and users on line $l_i$ are modeled by two independent HPPPs denoted by $\Psi_{l_i}^{b}$ with intensity $\lambda_b$ and  $\Psi_{l_i}^{u}$ with intensity $\lambda_u$ , respectively. Thus, the locations of the BSs and users are modeled by a PLC, which can be denoted by $\bar{\Phi}_b(\lambda_l,\lambda_b)$ and $\bar{\Phi}_u(\lambda_l,\lambda_u)$, respectively. Note that  $\bar{\Phi}_b(\lambda_l,\lambda_b)$ and $\bar{\Phi}_u(\lambda_l,\lambda_u)$ are driven by the same PLP.
For the ease of system {\color{black}illustration}, it is assumed that a user can only be served by the BSs from the same road. It is also assumed that all nodes considered in this paper are equipped with a single antenna.

Consider a downlink N-NOMA scenario in the considered vehicular network.
Specifically, for notational simplicity and according to Slivnyak's theorem \cite{haenggi2012stochastic}, we add an additional line $l_0$ which is coincided with the x-axis to the aforementioned PLP $\Phi_l$. Then a new line process $\Phi_{l_0}=\Phi_l\cup\{l_0\}$ is obtained.
A typical CoMP user belonging to $l_0$, is assumed to locate at the origin.
To make the application of CoMP meaningful, it is assumed that the distances from the CoMP user to the BSs on $l_0$ are larger than a given distance, which is denoted by $\mathcal{D}$. Then the BSs on $l_0$ are modeled by:
\begin{align}
\Psi_{l_0}^{b}=\{x_{L,m}^{l_0}\}\cup\{x_{R,m}^{l_0}\},m=1,2\cdots,
\end{align}
where $\{x_{L,m}^{l_0}\}$ {\color{black}denote} the points on the left hand side of the origin, which are modeled by a 1D HPPP on $(-\infty,\mathcal{D})$ with intensity $\lambda_b$,  and
$\{x_{R,m}^{l_0}\}$ denotes the points on the right hand side of the origin, which are modeled by a 1D HPPP on $(\mathcal{D},\infty)$ with intensity $\lambda_b$. It is worth pointing out that the points in $\{x_{t,m}^{l_0}\},t\in\{L,R\}$ are ordered by their distance to the origin, i.e., for $0<i\leq j$, $||x_{t,i}^{l_0}|| \leq ||x_{t,j}^{l_0}||$ ($t\in\{L,R\}$).

As shown in Fig. \ref{system_model}, the two BSs locating at $x_{L,1}^{l_0}$ and $x_{R,1}^{l_0}$ are scheduled to support the CoMP user cooperatively, while each BS is also individually serving a NOMA user by occupying the same resource block allocated to the CoMP user.

For notational simplicity, the CoMP user, the NOMA users associated with BSs locating at $x_{L,1}^{l_0}$ and $x_{R,1}^{l_0}$ are termed ``user $0$'', ``user $1$'' and ``user $2$'', respectively. The coordinate of user $k$ is denoted by $U_k$, $k=0,1,2$. It is assumed that the location of the NOMA user $k$ ($k=1,2$) is uniformly distributed in the segment on line $l_0$. Note that,  the segment is centered at the NOMA user's associating BS with length
$2\mathcal{R}$. It is also noteworthy that the NOMA user's distance to its serving BS {\color{black}is assumed to be} much smaller than that of the CoMP user.
\subsection{Channel Model}
The composite channel gain between user $k$ ($k=0,1,2$) and the $m$-th BS located on $l_i\in\Phi_l$ is modeled by:
\begin{align}	
\hat{h}_{i,m,k}=\frac{\hat{g}_{i,m,k}}{\sqrt{||x_{m}^{l_i}-U_k||^{\alpha_1}}},
\end{align}	
where $\hat{g}_{i,m,k}$ is the small scale Rayleigh fading gain,
i.e., $\hat{g}_{i,m,k}\sim CN(0,1)$, $x_{m}^{l_i}$ is the location of
the $m$-th BS located on $l_i\in\Phi_l$, $\alpha_1$ is the large scale path loss exponent.

Similarly, the composite channel gain between user $k$ ($k=0,1,2$) and a BS located at $x_{t,m}^{l_0}$ ($t\in\{L,R\}$) on $l_0$ is modeled by:
\begin{align}	
\tilde{h}_{t,m,k}=\frac{\tilde{g}_{t,m,k}}{\sqrt{||x_{t,m}^{l_0}-U_k||^{\alpha_0}}},
\end{align}	
where $\tilde{g}_{t,m,k}$ is the small scale Rayleigh fading gain,
i.e., $\tilde{g}_{t,m,k}\sim CN(0,1)$, {\color{black}and} $\alpha_0$ is the corresponding large scale path loss exponent.

It is worth pointing out that in the above channel modeling, different notations, i.e., $\alpha_0$ and $\alpha_1$, are used to denote the large scale path loss exponents. This is made to reflect the fact that the signal propagation condition between two nodes located on the same road differs from that between two nodes located on different roads.

\subsection{Description of N-NOMA}
In the considered N-NOMA scheme, the two considered cooperating BSs apply Alamouti coding in two time slots (termed ``time slot $1$'' and ``time slot $2$'') to serve the CoMP user. At the meantime, superposition coding is utilized by each of two BSs at each time slot to serve a NOMA user. It is assumed that the channel conditions remain unchanged over the two considered time slots. More specifically, the transmitted signals by the BS located at $x_{L,1}^{l_0}$ during time slots $1$ and $2$
are given by
\begin{align}\label{TxL1}
 s_{L,1}(1)=\bar{s}_{L,1}(1)+\tilde{s}_{c}(1),
\end{align}
and
\begin{align}\label{TxL2}
s_{L,1}(2)=\bar{s}_{L,1}(2)-\tilde{s}_{c}^*(2),
\end{align}
respectively, where $\bar{s}_{L,1}(1)$ and $\bar{s}_{L,1}(2)$ are the signals intended for
user $1$ at time slot $1$ and $2$, respectively, and $\tilde{s}_{c}(1)$ and $\tilde{s}_{c}(2)$ are signals intended for user $0$.
Similarly, the transmitted signals by the BS located at $x_{R,1}^{l_0}$ during time slots $1$ and $2$
are given by
\begin{align}\label{TxR1}
 s_{R,1}(1)=\bar{s}_{R,1}(1)+\tilde{s}_{c}(2),
\end{align}
and
\begin{align}\label{TxR2}
s_{R,1}(2)=\bar{s}_{R,1}(2)+\tilde{s}_{c}^*(1),
\end{align}
respectively.

Note that $\bar{s}_{p,1}(t)$ and $\tilde{s}_c(t)$ ($p\in\{L,R\}$, $t\in\{1,2\}$) are independently coded signals with Gaussian codebooks. The powers of these signals are given by
\begin{align}
 &\mathbb{E}\{|\bar{s}_{p,1}(t)|^2\}=\beta P,\\
 &\mathbb{E}\{|\tilde{s}_{c}(t)|^2\}=(1-\beta) P,
\end{align}
where $p\in\{L,R\}$ and $t\in\{1,2\}$, $P$ is the transmit power of a BS, and $\beta$ is the power allocation coefficient.

The observed signal at time slot $t$ ($t=1,2$) by user $k$ ($k=0,1,2$) is given by:
\begin{align}\label{rk}
 r_k(t)=\tilde{h}_{L,1,k}s_{L,1}(t)+\tilde{h}_{R,1,k}s_{R,1}(t)+I_k(t)+n_k(t),
\end{align}
where $n_k(t)$ denotes the additive noise, which is modeled as a symmetric complex Gaussian random variable, i.e., $n_k(t)\sim CN(0,\sigma^2)$, where $\sigma^2$ is the noise power. Note that, the noises for different $k$ and $t$ are not correlated.
$I_k(t)$ is the interference from other BSs, which can be expressed as follows:
\begin{align}\label{I_k(t)}
 I_k(t)=&\underset{I_{\text{intra},k}(t)}
                  {\underbrace{\sum_{j\in\{L,R\}}\sum_{m=2,3,\cdots}\tilde{h}_{j,m,k}s_{j,m}(t)}}
                  +\underset{I_{\text{inter},k}(t)}
                  {\underbrace{\sum_{l_i\in\Phi_l}\sum_{x_{m}^{l_i}\in{\Psi_{l_i}^b}}\hat{h}_{i,m,k}S_{i,m}(t)}}
\end{align}
where $s_{j,m}(t)$ is the signal transmitted by the $m$-th BS on the left/right hand side of the origin on line $l_0$,  and $S_{i,m}(t)$ is the signal transmitted by the $m$-th BS on line $l_i\in\Phi_l$. It is noteworthy that $I_{\text{intra},k}(t)$ is the interference from BSs on $l_0$ and hence is termed ``intra line interference'', and $I_{\text{inter},k}(t)$ is the interference from other roads and hence termed ``inter line interference''.

\section{{\color{black}Decoding Strategies and Signal-to-interference-plus-noise ratio (SINR) Modelling}}
This section discusses how to decode the signals at the CoMP and NOMA users. The expressions of the
SINRs are also provided, which are the preliminaries for the outage analysis.

 Note that, both {\color{black}the} CoMP and NOMA users need to decode the CoMP signal
 $\tilde{s}_c(1)$ and  $\tilde{s}_c(1)$  according to Alamouti scheme.
In order to implement the decoding, the received signal at user $k$ ($k=0,1,2$) is written as a vector denoted by
\begin{align}
 \mathbf{r}_k=[r_k(1),r_k^*(2)]^T.
\end{align}
Next, the detection matrix $\mathbf{H}_k$ is given by
\begin{align}
 \mathbf{H}_k=\frac{1}{\sqrt{|\tilde{h}_{L,1,k}|^2+|\tilde{h}_{R,1,k}|^2}} \left[
 \begin{matrix}
   \tilde{h}_{L,1,k} & \tilde{h}_{R,1,k} \\
   \tilde{h}_{R,1,k}^* & -\tilde{h}_{L,1,k}^*
  \end{matrix} \right].
\end{align}
By multiply $\mathbf{r}_k$ with $\mathbf{H}_k^{H}$, two parallel  single-input-single-output (SISO) channel expressions are obtained as follows:
\begin{align}\label{yk1}
 y_{k,1}=&C_k\tilde{s}_c(1)+
         \frac{|\tilde{h}_{L,1,k}|^2}{C_k}\bar{s}_{L,1}(1)
         +\frac{\tilde{h}_{L,1,k}^*\tilde{h}_{R,1,k}}{C_k}\bar{s}_{L,1}^*(2)
        +\frac{\tilde{h}_{L,1,k}^*\tilde{h}_{R,1,k}}{C_k}\bar{s}_{R,1}(1)\\\notag
        & +\frac{|\tilde{h}_{R,1,k}|^2}{C_k}\bar{s}_{R,1}^*(2)
         +\tilde{I}_{k,1}+\tilde{n}_{k,1}
\end{align}
\begin{align}\label{yk2}
 y_{k,2}=&C_k\tilde{s}_c(2)+
         \frac{\tilde{h}_{L,1,k}\tilde{h}_{R,1,k}^*}{C_k}\bar{s}_{L,1}(1)
         -\frac{|\tilde{h}_{L,1,k}|^2}{C_k}\bar{s}_{L,1}^*(2)
         +\frac{|\tilde{h}_{R,1,k}|^2}{C_k}\bar{s}_{R,1}(1)\\\notag
        & -\frac{\tilde{h}_{L,1,k}\tilde{h}_{R,1,k}^*}{C_k}\bar{s}_{R,1}^*(2)
         +\tilde{I}_{k,2}+
         \tilde{n}_{k,2}
\end{align}

where $C_k=\sqrt{|\tilde{h}_{L,1,k}|^2+|\tilde{h}_{R,1,k}|^2}$,
\begin{align}
 	 \tilde{n}_{k,p}=\theta_{k,p,1}n_k(1)+\theta_{k,p,2}n_k^*(2), p\in\{1,2\}	,
\end{align}
and
\begin{align}
 \tilde{I}_{k,p}=\theta_{k,p,1}I_k(1)+\theta_{k,p,2}I_k^*(2), p\in\{1,2\}	,
\end{align}
and
\begin{align}
 \left(\theta_{k,1,1},\theta_{k,1,2}\right)
 =\left(\frac{\tilde{h}_{L,1,k}^*}{C_k},
         \frac{\tilde{h}_{R,1,k}}{C_k}\right),
\end{align}
and
\begin{align}
 \left(\theta_{k,2,1},\theta_{k,2,2}\right)
 =\left(\frac{\tilde{h}_{R,1,k}^*}{C_k},
         -\frac{\tilde{h}_{L,1,k}}{C_k}\right).
\end{align}

Based on (\ref{yk1}) and (\ref{yk2}), the CoMP user's signal $\tilde{s}_c(1)$ and $\tilde{s}_c(2)$ can be decoded separately through $y_{k,1}$ and $y_{k,2}$ at user $k$, respectively.
Further, to evaluate the performance of the proposed scheme, it is necessary to first characterize the power of  $\tilde{n}_{k,p}$ and $\tilde{I}_{k,p}$.
It can be easily concluded that $\tilde{n}_{k,p}$ ($p\in\{1,2\}$) has the same distribution
as $n_k(t)$, i.e., $\tilde{n}_{k,p} \sim CN(0,\sigma^2)$, since $|\theta_{k,p,1}|^2+|\theta_{k,p,2}|^2=1$ and $n_k(1)$ and $n_k(2)$ are i.i.d Gaussian distributed.
However, the situation for $\tilde{I}_{k,p}$ is more challenging, which will be discussed in the following subsection.
\subsection{{\color{black}Modelling} $\tilde{I}_{k,p}$}
Note that if we assume all the interfering BSs work individually without cooperating with other BSs and transmit independent signals over different time slots, then the expected
power of the interference can be expressed as follows:
\begin{align}\label{inter_power}
P_{I_k}=&\mathbb{E}\{|\tilde{I}_{k,p}|^2\}\\\notag
=&P\left(\sum_{j\in\{L,R\}}\sum_{m=2,3,\cdots}\left|\tilde{h}_{j,m,k}\right|^2
   +\sum_{l_i\in\Phi_l}\sum_{x_{m}^{l_i}\in{\Psi_{l_i}^b}}\left|\hat{h}_{i,m,k}\right|^2\right)\\\notag
\overset{\Delta}{=}&P \zeta_k,
\end{align}
where $p\in\{1,2\}$.

However, in this paper, the above assumption does not make sense in the considered large scale network. The reason is as follows. Since the proposed N-NOMA is introduced into the whole network, it is not reasonable to assume that only one pair of BSs, i.e., the BSs locating at $x_{L,1}^{l_0}$ and $x_{R,1}^{l_0}$, working cooperatively by applying the proposed N-NOMA, whereas all other BSs in the network work individually without cooperation. A practical assumption for the considered network will {\color{black}be stated} as follows:
\begin{Assumption}
According to whether adopting the proposed N-NOMA scheme, the interfering BS in the  network can be classified into two disjoint sets as follows:
\begin{align}
\Psi_{l_0}^b\backslash \{x_{L,1}^{l_0},x_{R,1}^{l_0}\}
 \cup\underset{l_i\in\Phi_l}{\cup}\Psi_{l_i}^b=\Omega_1\cup\Omega_2.
\end{align}
where
 \begin{itemize}
   \item in $\Omega_1$, each BS works individually and transmits independent signals over different time slots.
   \item in $\Omega_2$, all BSs work in pairs cooperatively by applying the proposed
   N-NOMA scheme.
 \end{itemize}
\end{Assumption}

Note that, based on Assumption $1$, the interferences observed at time slots $1$ and $2$,
i.e., $I_k(1)$ and $I_k(2)$, are correlated with each other, because the signals intended for
the CoMP users are transmitted over the two time slots by the corresponding cooperating BSs in pairs.  Intuitively, it seems that the above correlation will significantly complicate the analysis for the interference. However, {\color{black}surprisingly}, through rigorous derivations, it is shown that in terms of the expected  power of the aggregated interference $\tilde{I}_{k,p}$, Assumption $1$ is equivalent to the assumption that all interfering BS work individually, as highlighted in the following lemma.

\begin{Lemma}
Based on Assumption $1$, the expected  power of the aggregated interference $\tilde{I}_{k,p}$
($p\in\{1,2\}$), can be expressed as shown in  (\ref{inter_power}).
\end{Lemma}
\begin{IEEEproof}
Please refer to Appendix A.
\end{IEEEproof}
\subsection{{\color{black}Modelling the SINR}}
According to NOMA principle, user $0$, i.e., the CoMP user, decodes its  signal
$\tilde{s}_c(1)$ and $\tilde{s}_c(2)$ with the following SINR:
\begin{align}
 \text{SINR}_{0}=\frac{C_0^2(1-\beta)}{C_0^2\beta +\zeta_0
                                              +1/\rho},
\end{align}
where the signals intended for the NOMA users, i.e., $\bar{s}_{p,1}(t)$ ($p\in\{1,2\},t\in\{1,2\}$)£¬ are treated as additive noise, and $\rho = P/\sigma^2$
is the transmit signal-to-noise ratio (SNR).

{\color{black}User} $j$ ($j\in\{1,2\}$), i.e., the NOMA user, firsts decode the CoMP user's signal $\tilde{s}_c(1)$ and $\tilde{s}_c(2)$ from (\ref{yk1}) and (\ref{yk2}) with
\begin{align}
\text{SINR}_{j,0}=\frac{C_j^2(1-\beta)}{C_j^2\beta +\zeta_j
                                              +1/\rho},
\end{align}
If successful, user $j$ ($j\in\{1,2\}$) carries out successive interference cancellation to remove the CoMP user's signal in (\ref{rk}). Then, the modified observation at user $j$ can be expressed as:
\begin{align}\label{trk}
 \tilde{r}_j(t)=\tilde{h}_{L,1,j}\bar{s}_{L,1}(t)+\tilde{h}_{R,1,j}\bar{s}_{R,1}(t)+I_k(t)+n_k(t)
\end{align}
Finally, user $j$ ($j\in\{1,2\}$) decodes its own signal from (\ref{trk}) with
\begin{align}
 \text{SINR}_{1,1}=\frac{|\tilde{h}_{L,1,1}|^2\beta }{|\tilde{h}_{R,1,1}|^2\beta +\zeta_1
                                              +1/\rho},
\end{align}
for user $1$ and
\begin{align}
 \text{SINR}_{2,1}=\frac{|\tilde{h}_{R,1,2}|^2\beta }{|\tilde{h}_{L,1,2}|^2\beta +\zeta_2
                                              +1/\rho},
\end{align}
for user $2$, respectively.

\section{Performance Analysis}
In this section, the outage probability is used as a criterion to characterize the performance of the proposed N-NOMA by using stochastic {\color{black}geometry}.
\subsection{CoMP user}
The outage probability achieved by the CoMP user (user $0$) is given by
\begin{align}
 P_0^{out}=P\left(\text{SINR}_0<\epsilon_0\right)
\end{align}
where $\epsilon_k=2^{R_k}-1$, $R_k$ is the target rate of user $k$, $k\in\{0,1,2\}$.

To obtain the expression for $ P_0^{out}$,  it is necessary to
characterize the Laplace transform of the normalized interference power $\zeta_0$ observed at the
CoMP user.
Note that $\zeta_0$ can be divided into  two independent components:
(i) the interference from the nodes on the typical line $l_0$, which can be denoted by
$\zeta_0^{\text{intra}}=\sum_{j\in\{L,R\}}\sum_{m=2,3,\cdots}\left|\tilde{h}_{j,m,0}\right|^2$;
(ii) the interference from the nodes on lines other than $l_0$, which can be denoted by
$\zeta_0^{\text{inter}}=\sum_{l_i\in\Phi_l}\sum_{x_{m}^{l_i}\in{\Psi_{l_i}^b}}\left|\hat{h}_{i,m,0}\right|^2$.  The Laplace transforms of $\zeta_0^{\text{intra}}$ and $\zeta_0^{\text{inter}}$ are highlighted as in the following.

Denote the distances from the CoMP user to the two serving BSs
locating at $x_{L,1}^{l_0}$ and $x_{R,1}^{l_0}$ by $d_1$ and $d_2$, i.e., $d_1=||x_{L,1}^{l_0}||$ and  $d_2=||x_{R,1}^{l_0}||$. By assuming $d_1$ and $d_2$ are fixed, we have the  following lemma which characterizes the intra line interference.

\begin{Lemma}
Given the distances from the CoMP user to the two serving BSs, i.e., $d_1$ and $d_2$, the conditional Laplace transform of the interference from the nodes on the typical line $l_0$ is given by
\begin{align}\label{Lap_intra_0}
 \mathcal{L}^{\text{intra}}(s,d_1,d_2)=  \exp\bigg(-\lambda_b
  \bigg(
         &\frac{sd_1^{1-\alpha_0}}{\alpha_0-1}{}_2F_1\left(1,1-\frac{1}{\alpha_0};2-\frac{1}{\alpha_0};-\frac{s}{d_1^{\alpha_0}}\right)\\\notag
         &+
         \frac{sd_2^{1-\alpha_0}}{\alpha_0-1}{}_2F_1\left(1,1-\frac{1}{\alpha_0};2-\frac{1}{\alpha_0};-\frac{s}{d_2^{\alpha_0}}\right)
  \bigg)\bigg),
\end{align}
where ${}_2F_1(\cdot)$ is the Gauss hyper-geometric function.
\end{Lemma}
\begin{IEEEproof}
Please refer to Appendix B.
\end{IEEEproof}

The Laplace transform of the interference from nodes on lines other than $l_0$ is given in the following lemma.
\begin{Lemma}
The Laplace transform of the interference from the nodes on lines other than $l_0$ is given by
\begin{align}
\mathcal{L}^{\text{inter}}(s)=\exp\left(-2\pi\lambda_l\int_{0}^{\infty}\left(1-G(x,s)\right)\,dx\right),
\end{align}
where
\begin{align}
G(x,s)=\exp\left(-2\lambda_b\int_{0}^{\infty}\frac{s}{s+(x^2+u^2)^\frac{\alpha_1}{2}}\,du\right).
\end{align}
\end{Lemma}
\begin{IEEEproof}
Please refer to Appendix C.
\end{IEEEproof}

 {\color{black}Because} $\zeta_0^{\text{intra}}$ and $\zeta_0^{\text{inter}}$ are independent of each other and $\zeta_0=\zeta_0^{\text{intra}}+\zeta_0^{\text{inter}}$, the Laplace transform of $\zeta_0$ can be easily expressed as
\begin{align}
 \mathcal{L}(s,d_1,d_2)=\mathcal{L}^{\text{intra}}(s,d_1,d_2)
                        \mathcal{L}^{\text{inter}}(s).
\end{align}

Based on the above preliminary results,
the outage probability achieved by the CoMP user
can be obtained as presented in the following theorem.
\begin{theorem}
Given the distances from the CoMP user to the two serving BSs, i.e., $d_1$ and $d_2$, the conditional outage probability achieved by the CoMP user can be expressed as
follows:
\begin{itemize}
	\item when $d_1\neq d_2$
\begin{align}\label{P1_ne}
P_0^{out}(d_1,d_2)=&1-\frac{d_2^{\alpha_0}}{d_2^{\alpha_0}-d_1^{\alpha_0}}
             e^{-\frac{\mu(\epsilon_0,d_1)}{\rho}}
             \mathcal{L}\left(\mu(\epsilon_0,d_1),d_1,d_2\right)\\\notag
          &+\frac{d_1^{\alpha_0}}{d_2^{\alpha_0}-d_1^{\alpha_0}}
             e^{-\frac{\mu(\epsilon_0,d_2)}{\rho}}
             \mathcal{L}\left(\mu(\epsilon_0,d_2),d_1,d_2\right),
\end{align}
\item when $d_1=d_2$
\begin{align}\label{P1_e}
	   P_0^{out}(d_1,d_2)=&1-e^{-\frac{\mu(\epsilon_0,d_1)}{\rho}}\left(1+\frac{\mu(\epsilon_0,d_1)}{\rho}\right)
	        \mathcal{L}(\mu(\epsilon_0,d_1),d_1,d_1)\\\notag
	    &\quad+e^{-\frac{\mu(\epsilon_0,d_1)}{\rho}}\mu(\epsilon_0,d_1)
	     \mathcal{L}^{(1)}(\mu(\epsilon_0,d_1),d_1,d_1).
	 \end{align}
\end{itemize}
where $\mu(\epsilon,d)=\frac{d^{\alpha_0}\epsilon}{1-\beta-\epsilon\beta}$, $\mathcal{L}^{(1)}(\mu(\epsilon_0,d_1),d_1,d_1)$ is the derivative of $\mathcal{L}(\mu(\epsilon_0,d_1),d_1,d_1)$ with respect to $\mu(\epsilon_0,d_1)$.
\end{theorem}
\begin{IEEEproof}
Please refer to Appendix D.
\end{IEEEproof}
\subsection{NOMA user}
Note that, the NOMA user's message can be successfully decoded only when the following two conditions are met i) the NOMA user can decode the message intended for the CoMP user, ii) the NOMA user can decode its own message after removing the CoMP user's message. Thus the outage probability of the NOMA user $k$ ($k=1,2$) can be expressed as
\begin{align}\label{Pk_out_1}
 P_k^{out}=1-P\left(\text{SINR}_{k,0}>\epsilon_0, \text{SINR}_{k,1}>\epsilon_1\right).
\end{align}
After some rigorous derivations, it is found that the expression for the outage probability achieved by the NOMA user $k$ ($k=1,2$) can be divided into two cases, which is dependent on the relationship among the power allocation coefficient $\beta$, the SINR thresholds $\epsilon_0$ of the CoMP user and the SINR threshold $\epsilon_k$ of the NOMA user $k$ ($k=1,2$).
The following theorem characterizes the outage performance achieved by the NOMA user.
\begin{theorem}
Given $d_1$ and $d_2$, the outage probability achieved by the NOMA user $k$ ($k=1,2$) can be expressed as the following two cases:
\begin{itemize}
\item when $\epsilon_k\geq \frac{\epsilon_0\beta}{1-\beta-\epsilon_0\beta}$
\begin{align}
 P_k^{out}(d_1,d_2)\approx&1-\frac{\pi}{4N}\sum\limits_{n=1}^{N}\sum\limits_{t=1}^2\sqrt{1-\theta_n^2}
                    \left(\frac{(d_1+d_2+(-1)^tc_n)^{\alpha_0}}{(d_1+d_2+(-1)^tc_n)^{\alpha_0}+\epsilon_k
                    c_n^{\alpha_0}}\right.\\\notag
                    &\quad\quad\quad\left.\exp\left(-\frac{c_n^{\alpha_0}\epsilon_k}{\beta\rho}\right)\mathcal{L}_{1,t}(\frac{c_n^{\alpha_0}\epsilon_k}{\beta},d_1,d_2,c_n)\right),
\end{align}
\item when $\epsilon_k < \frac{\epsilon_0\beta}{1-\beta-\epsilon_0\beta}$
\begin{align}
 P_k^{out}(d_1,d_2)\approx1-\frac{\pi}{4N}\sum\limits_{n=1}^{N}\sqrt{1-\theta_n^2}(A_{k,n,1}+A_{k,n,2}+D_{k,n,1}+D_{k,n,2}),
\end{align}
\end{itemize}
where $N$ is parameter of the Gauss-Chebyshev quadrature, $\theta_n=\cos{\frac{(2n-1)\pi}{2N}}$, $c_n=\frac{\mathcal{R}}{2}(1+\theta_n)$, $\mathcal{L}_{1,1}(\cdot)$ and  $\mathcal{L}_{1,2}(\cdot)$ are given by (\ref{Lap_NOMA11}) and (\ref{Lap_NOMA12}) as shown in Appendix E, and
\begin{align}
&A_{k,n,t}=\frac{(d_1+d_2+(-1)^{t}c_n)^{\alpha_0}}{(d_1+d_2+(-1)^{t}c_n)^{\alpha_0}-c_n^{\alpha_0}}
        \bigg(e^{-\frac{c_n^{\alpha_0}\epsilon_0}{(1-\beta-\epsilon_0\beta)\rho}}
             \mathcal{L}_{1,t}(\frac{c_n^{\alpha_0}\epsilon_0}{(1-\beta-\epsilon_0\beta)},d_1,d_2,c_n)\\\notag
             &\quad\quad\quad\quad\quad\quad\quad\quad\quad\quad\quad\quad
             -e^{-\phi_{k,n,t}/\rho}\mathcal{L}_{1,t}(\phi_{k,n,t},d_1,d_2,c_n)\bigg),\\\notag
 &D_{k,n,t}=\frac{(d_1+d_2+(-1)^{t}c_n)^{\alpha_0}}{(d_1+d_2+(-1)^{t}c_n)^{\alpha_0}+\epsilon_kc_n^{\alpha_0}}
         e^{-\psi_{k,n,t}/\rho}\mathcal{L}_{1,t}(\psi_{k,n,t},d_1,d_2,c_n),\\\notag
 &\phi_{k,n,t}=((d_1+d_2+(-1)^tc_n)^{\alpha_0}-c_n^{\alpha_0})\eta_k+\frac{c_n^{\alpha_0}\epsilon_0}{1-\beta-\epsilon_0\beta},\\\notag
 &\psi_{k,n,t}=\frac{c_n^{\alpha_0}\epsilon_k}{\beta}+\left((d_1+d_2+(-1)^tc_n)^{\alpha_0}+c_n^{\alpha_0}\epsilon_k\right)\eta_k,\\\notag
 &\eta_k=\frac{\epsilon_0\beta-\epsilon_k(1-\beta-\epsilon_0\beta)}
	                {\beta(1-\beta-\epsilon_0\beta)(1+\epsilon_k)}.
\end{align}
\end{theorem}
\begin{IEEEproof}
Please refer to Appendix E.
\end{IEEEproof}

\subsection{Asymptotic analysis}
In this subsection, we {\color{black}focus on} the case {\color{black}in which} the intensity of the roads $\lambda_l$ becomes extremely large and the intensity of the BSs on each road $\lambda_b$ becomes extremely low while the product of $\lambda_l$ and $\lambda_b$ remains constant. The outage probabilities achieved by the CoMP and NOMA users under this limit case are given by the following two corollaries.
\begin{Corollary}
When $\lambda_l\to \infty$, $\lambda_b \to 0$, and $\lambda_l\lambda_b=\lambda$, the outage probability achieved by the CoMP user can be approximated as follows:
\begin{itemize}
	\item when $d_1\neq d_2$
\begin{align}
P_0^{out}(d_1,d_2)\approx&1-\frac{d_2^{\alpha_0}}{d_2^{\alpha_0}-d_1^{\alpha_0}}
             e^{-\frac{\mu(\epsilon_0,d_1)}{\rho}}
             \bar{\mathcal{L}}\left(\mu(\epsilon_0,d_1)\right)\\\notag
          &+\frac{d_1^{\alpha_0}}{d_2^{\alpha_0}-d_1^{\alpha_0}}
             e^{-\frac{\mu(\epsilon_0,d_2)}{\rho}}
             \bar{\mathcal{L}}\left(\mu(\epsilon_0,d_2)\right),
\end{align}
\item when $d_1=d_2$
\begin{align}
	   P_0^{out}(d_1,d_2)\approx&1-e^{-\frac{\mu(\epsilon_0,d_1)}{\rho}}\left(1+\frac{\mu(\epsilon_0,d_1)}{\rho}\right)
	       \bar{\mathcal{L}}(\mu(\epsilon_0,d_1))\\\notag
	    &\quad+e^{-\frac{\mu(\epsilon_0,d_1)}{\rho}}\mu(\epsilon_0,d_1)
	     \bar{\mathcal{L}}^{(1)}(\mu(\epsilon_0,d_1)).
	 \end{align}
\end{itemize}
where
\begin{align}\label{Lap_limit}
\bar{\mathcal{L}}=\exp\left(-\frac{2\pi^2\lambda s^{\frac{2}{\alpha_1}}}{\alpha_1}B\left(\frac{2}{\alpha_1},1-\frac{2}{\alpha_1}\right)\right),
\end{align}
$B(\cdot,\cdot)$ is the beta function, and $\bar{\mathcal{L}}^{(1)}(s)$ is the first derivative of $\bar{\mathcal{L}}(s)$.
\end{Corollary}

\begin{Corollary}
When $\lambda_l\to \infty$, $\lambda_b \to 0$, and $\lambda_l\lambda_b=\lambda$, the outage probability achieved by the NOMA user $k$ ($k=1,2$) can be approximated as follows:
\begin{itemize}
\item when $\epsilon_k\geq \frac{\epsilon_0\beta}{1-\beta-\epsilon_0\beta}$
\begin{align}
 P_k^{out}(d_1,d_2)\approx&1-\frac{\pi}{4N}\sum\limits_{n=1}^{N}\sum\limits_{t=1}^2\sqrt{1-\theta_n^2}
                    \left(\frac{(d_1+d_2+(-1)^tc_n)^{\alpha_0}}{(d_1+d_2+(-1)^tc_n)^{\alpha_0}+\epsilon_k
                    c_n^{\alpha_0}}\right.\\\notag
                    &\quad\quad\quad\left.\exp\left(-\frac{c_n^{\alpha_0}\epsilon_k}{\beta\rho}\right)
                   \bar{\mathcal{L}}(\frac{c_n^{\alpha_0}\epsilon_k}{\beta})\right),
\end{align}
\item when $\epsilon_k < \frac{\epsilon_0\beta}{1-\beta-\epsilon_0\beta}$
\begin{align}
 P_k^{out}(d_1,d_2)\approx1-\frac{\pi}{4N}\sum\limits_{n=1}^{N}\sqrt{1-\theta_n^2}(\bar{A}_{k,n,1}+\bar{A}_{k,n,2}+\bar{D}_{k,n,1}+\bar{D}_{k,n,2}),
\end{align}
\end{itemize}
where
\begin{align}
&\bar{A}_{k,n,t}=\frac{(d_1+d_2+(-1)^{t}c_n)^{\alpha_0}}{(d_1+d_2+(-1)^{t}c_n)^{\alpha_0}-c_n^{\alpha_0}}
        \bigg(e^{-\frac{c_n^{\alpha_0}\epsilon_0}{(1-\beta-\epsilon_0\beta)\rho}}
             \bar{\mathcal{L}}(\frac{c_n^{\alpha_0}\epsilon_0}{(1-\beta-\epsilon_0\beta)})\\\notag
             &\quad\quad\quad\quad\quad\quad\quad\quad\quad\quad\quad\quad
             -e^{-\phi_{k,n,t}/\rho}\bar{\mathcal{L}}(\phi_{k,n,t})\bigg),\\\notag
 &\bar{D}_{k,n,t}=\frac{(d_1+d_2+(-1)^{t}c_n)^{\alpha_0}}{(d_1+d_2+(-1)^{t}c_n)^{\alpha_0}+\epsilon_kc_n^{\alpha_0}}
         e^{-\psi_{k,n,t}/\rho}\bar{\mathcal{L}}(\psi_{k,n,t}).
\end{align}
\end{Corollary}

It is {\color{black}worth pointing out } that the expression (\ref{Lap_limit}) for the Laplace transform of the interference in the considered limit case is the same as that of the interference in the conventional 2D Poisson bipolar networks \cite{haenggi2012stochastic,haenggi2015meta}, where the intensity density of the 2D
HPPP is $\pi\lambda$. Hence, it indicates that the locations of the interfering BSs of the PLC
 under the limit case degrade to a 2D HPPP.

\section{Numerical results}
In this section, numerical results are presented to evaluate the performance achieved by the proposed N-NOMA scheme and also verify the accuracy of the developed analytical results. Unless stated otherwise, the parameters are set as follows: the thermal noise is $-170$ dBm/Hz, the carrier frequency is $2 \times 10^9$ Hz, the channel bandwidth is $10$ MHz, the transmission power is $30$ dBm, $\lambda_l=5\times10^{-4}$ $\text{nodes}/m^2$, $\lambda_b=5\times10^{-3}$ $\text{nodes}/m$, $\alpha_0=3$.
\begin{figure}[!t]
\centering
\includegraphics[width=3.5in]{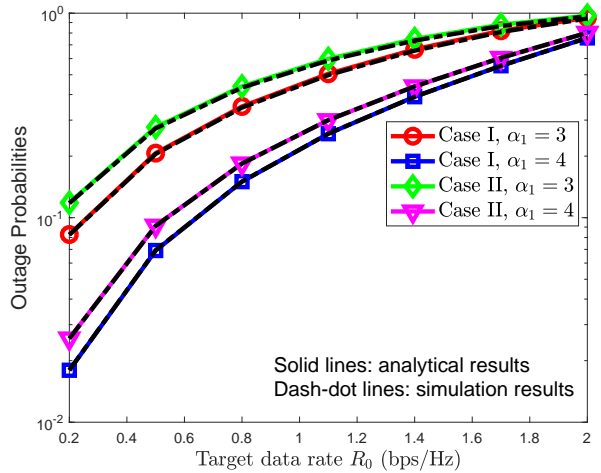}
\caption{CoMP user's outage probability. Case I: $d_1=100$ m, $d_2=100$ m; Case II: $d_1=100$ m, $d_2=150$ m.  $\beta=\frac{1}{5}$.}
\label{accuracy_CoMP}
\end{figure}

\begin{figure}[!t]
\centering
\includegraphics[width=3.5in]{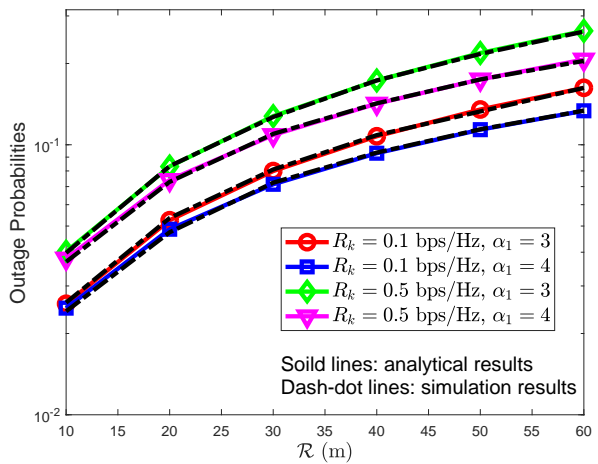}
\caption{NOMA user's outage probability. $d_1=d_2=100$ m, $\beta=\frac{1}{5}$, $R_0=0.5$ bps/Hz.}
\label{accuracy_NOMA}
\end{figure}

Figs. \ref{accuracy_CoMP} and \ref{accuracy_NOMA} show the outage probabilities achieved by the CoMP and NOMA users, respectively. The analytical results in Fig. \ref{accuracy_CoMP} are based on Theorem $1$, while {\color{black}those} in Fig. \ref{accuracy_NOMA} are based on Theorem $2$. The simulation results are obtained by generating $100000$ realizations of the PLC. As shown in Figs. \ref{accuracy_CoMP} and \ref{accuracy_NOMA}, the simulation results perfectly matches the theoretical results, which verifies the accuracy the developed analysis.

\begin{figure*}[!t]
\centering
\subfloat[Outage sum rates]{\includegraphics[width=3.2in]{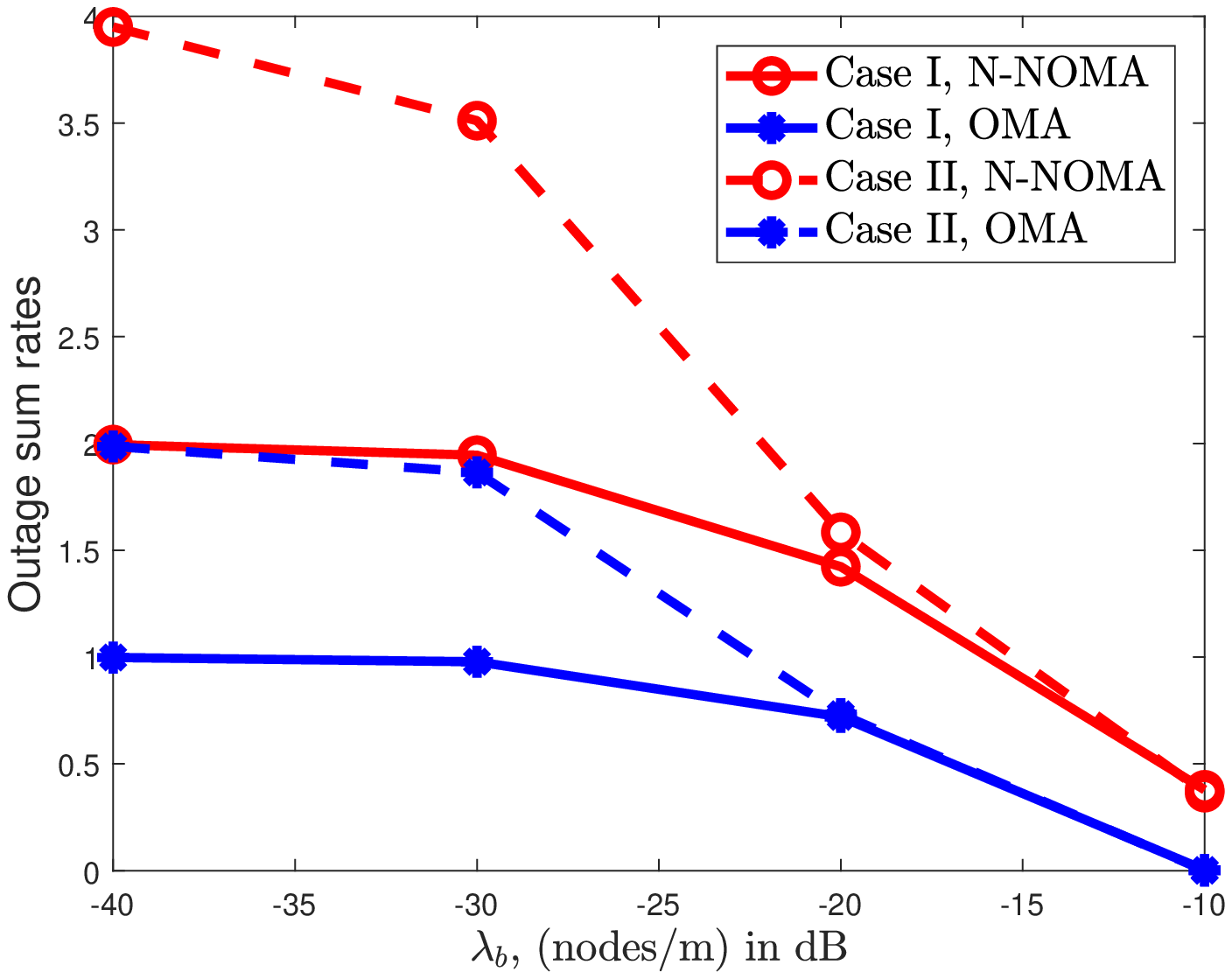}%
\label{compare_rate}}
\hfil
\subfloat[Outage probabilities]{\includegraphics[width=3.2in]{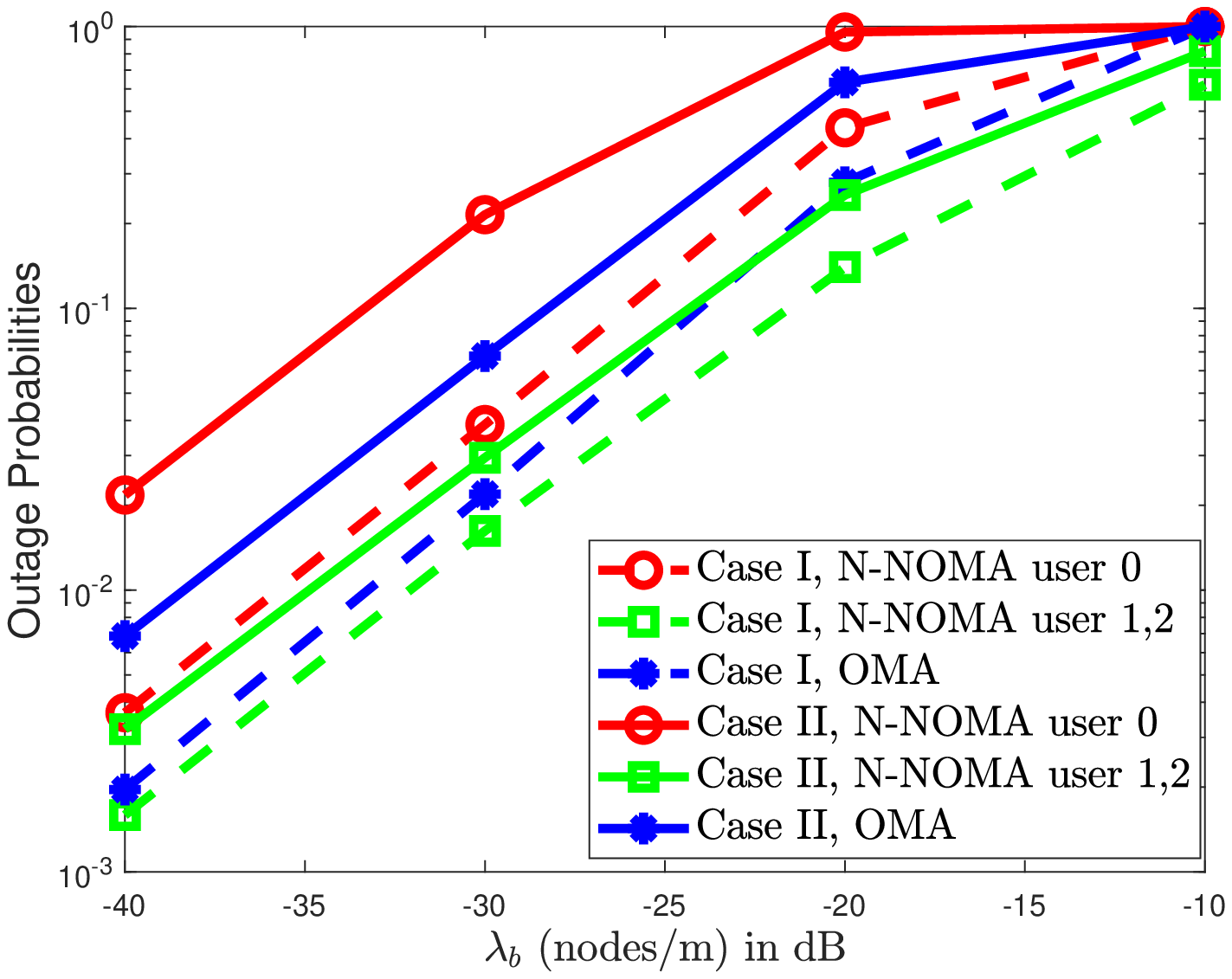}%
\label{compare_outage}}
\caption{Performance comparison of N-NOMA and conventional OMA.
Case I: $R_0=1$ bps/Hz, $R_k=0.5$ bps/Hz ($k=1,2$);
Case II: $R_0=2$ bps/Hz, $R_k=1$ bps/Hz ($k=1,2$).
$d_1=d_2=100$ m, $\alpha_1=4$, $\beta=\frac{1}{5}$, $\mathcal{R}=20$ m.}
\label{compare}
\end{figure*}

Fig. \ref{compare} studies the performance comparison of the proposed NOMA scheme and the conventional OMA scheme. The outage sum rates are shown in Fig. \ref{compare}(a), while the corresponding outage probabilities are shown in Fig. \ref{compare}(b). It is worth pointing out that, in the benchmark OMA scheme, only the CoMP user is served by the two BSs by applying the Alamouti code. As shown in Fig. \ref{compare}(a), the proposed N-NOMA scheme outperforms the conventional OMA scheme in terms of outage sum rates. For example, in Case II, when $\lambda_b=10^{-3}$ nodes/m, the sum rate achieved by the N-NOMA scheme is about $3.5$ bps/Hz, while that of the OMA scheme is about $2$ bps/Hz. Hence, the N-NOMA scheme {\color{black}realizes a gain} of $1.5$  bps/Hz over the OMA scheme.

\begin{figure*}[!t]
\centering
\subfloat[Outage probabilities]{\includegraphics[width=3.2in]{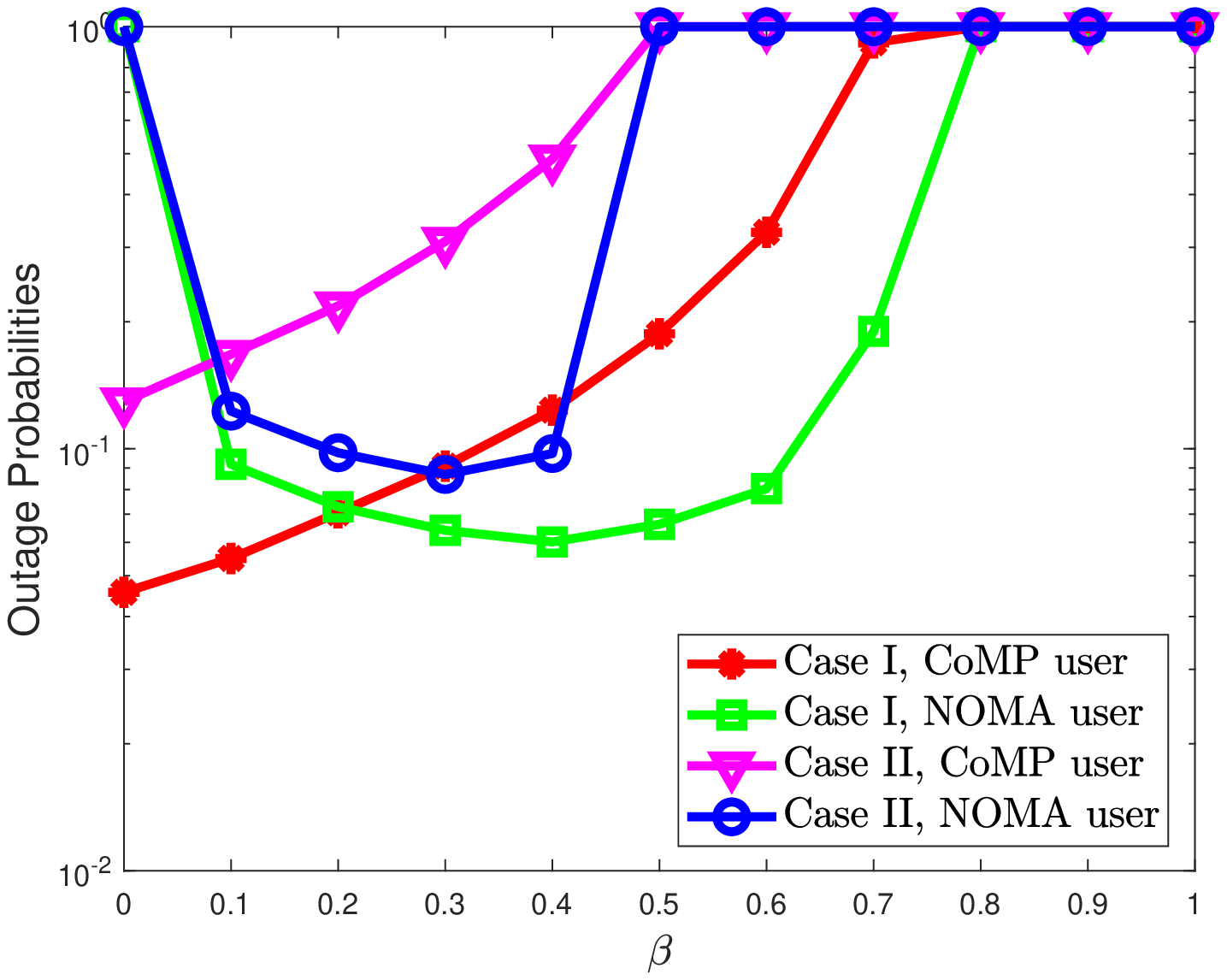}%
\label{impact_beta_outage}}
\hfil
\subfloat[Outage sum rates]{\includegraphics[width=3.2in]{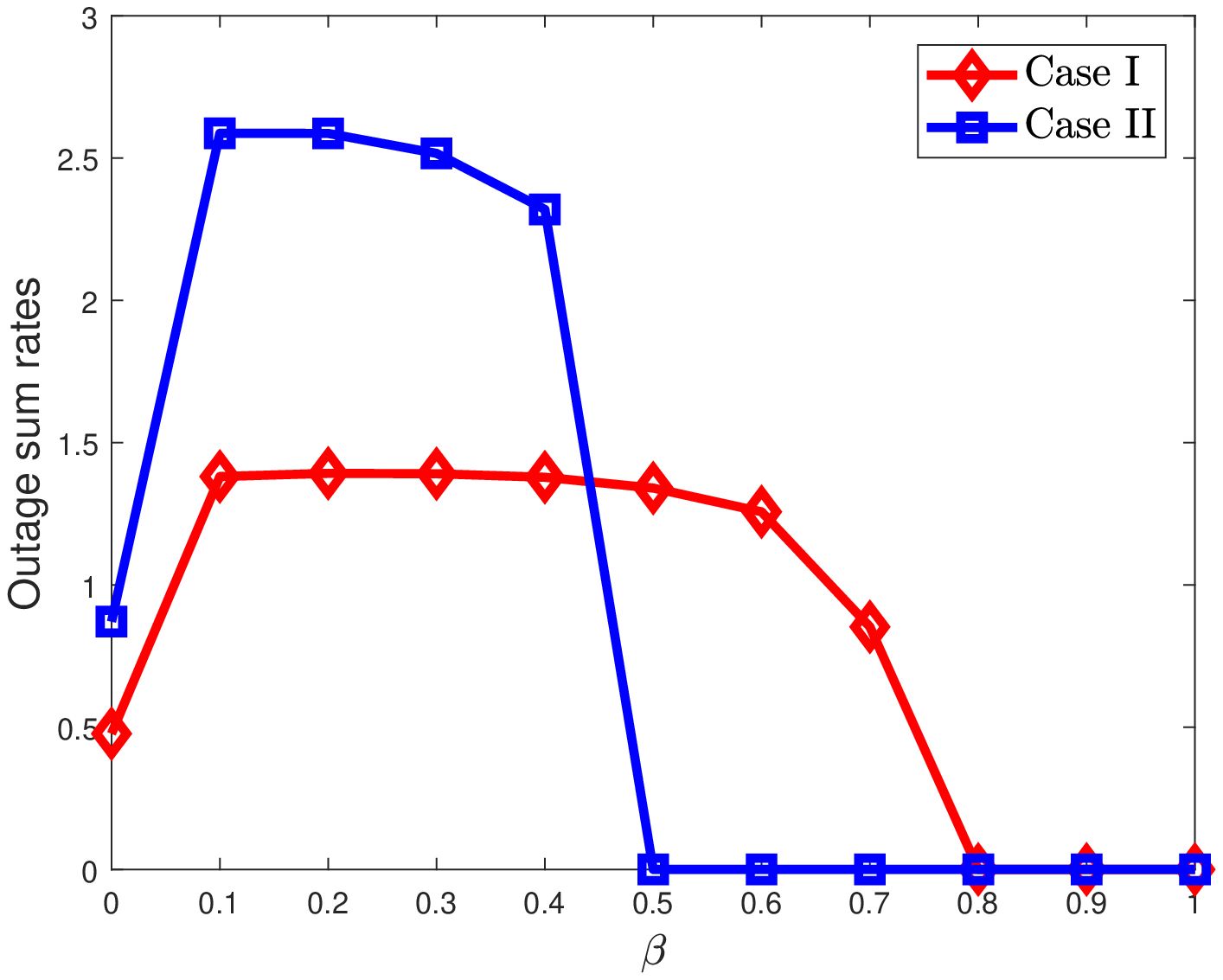}%
\label{impact_beta_rate}}
\caption{Impact of the power allocation coefficient $\beta$ on the performance.
 Case I: $R_0=0.5$ bps/Hz, $R_k=0.5$ bps/Hz ($k=1,2$);
Case II: $R_0=1$ bps/Hz, $R_k=1$ bps/Hz ($k=1,2$).
$d_1=d_2=100$ m, $\alpha_1=4$, $\mathcal{R}=20$ m.}
\label{impact_beta}
\end{figure*}

Fig. \ref{impact_beta} studies the impact of the power allocation coefficient $\beta$ on the performance, where the impact on the outage probabilities is shown in Fig. \ref{impact_beta}(a) and the  impact on the outage sum rates is shown in Fig. \ref{impact_beta}(b).
It is shown in Fig. 4(a) that the outage probability achieved by the CoMP user increases with $\beta$. This is because the CoMP user treats the NOMA users' signals as interference. Hence, increasing $\beta$ means more power is allocated to the interference and less power is allocated to the useful signal.
It is also shown in Fig. \ref{impact_beta}(a) that  the outage probability achieved by the NOMA user first decreases with $\beta$ and then increases. This can be explained as follows.
Note that the NOMA user needs to first carry out SIC to remove the CoMP user's signal, and then decode its own signal. When $\beta$ is small, it is with very high probability that the CoMP user's signal is successfully removed, and hence, decoding the NOMA user's own signal is the main limitation of the outage probability. However, as $\beta$ increases, removing the CoMP user's signal becomes the main limitation.
Fig. \ref{impact_beta}(b) shows that the outage sum rates first increases with $\beta$, {\color{black}and then decreases to $0$}.

\begin{figure}[!t]
\centering
\includegraphics[width=3.5in]{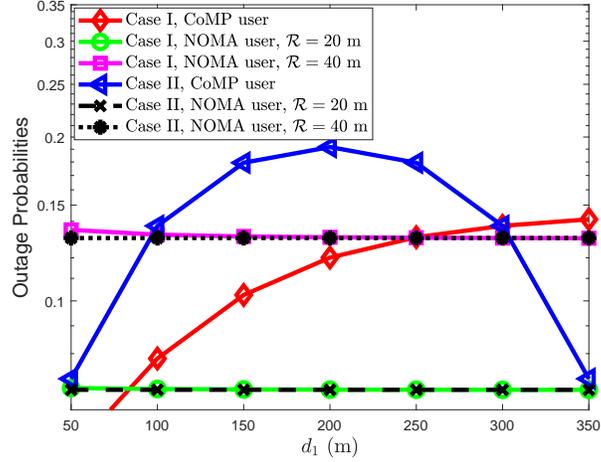}
\caption{Impact of the distances between the serving BSs and the CoMP user on the performance.
Case I: $d_2=100$ m; Case II: $d_1+d_2=400$ m.
$\alpha_1=4$, $\beta=\frac{1}{4}$, $R_k=0.5$ bps/Hz (k=0,1,2).}
\label{impact_d}
\end{figure}

Fig. \ref{impact_d} studies the impact of the distances $d_1$ and $d_2$ on the outage probabilities. Particularly, two cases are considered. In Case I, $d_2$ is fixed {\color{black}as} $100$ m , while the sum of $d_1$ and $d_2$ is fixed as $400$ m  in Case II. Interestingly, as shown in the figure the impacts of $d_1$ and $d_2$ on the NOMA user's outage probability {\color{black}are} negligible. This can be explained by the fact that the distance of two serving BSs is far {\color{black}such that} one BS has {\color{black}an} negligible effect on the other BS's NOMA user. It can also be shown in the figure that, the outage probability achieved by the CoMP user increases with $d_1$ in Case I, while that first increases with $d_1$ and then decreases in Case II.  Another interesting observation as shown in Fig. \ref{impact_d} is that the optimal point in Case II is that $d_1=d_2=200$ m.

\begin{figure}[!t]
\centering
\includegraphics[width=3.5in]{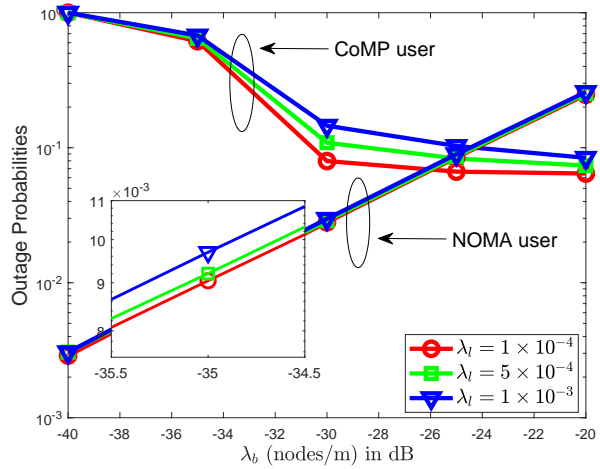}
\caption{Impact of $\lambda_l$ and $\lambda_b$ on the performance. $\alpha_1=4$,
$\beta=\frac{1}{4}$, $\mathcal{R}=40$ m, $R_k=0.5$ bps/Hz (k=0,1,2).}
\label{impact_lambda1}
\end{figure}

\begin{figure*}[!t]
\centering
\subfloat[CoMP user]{\includegraphics[width=3.2in]{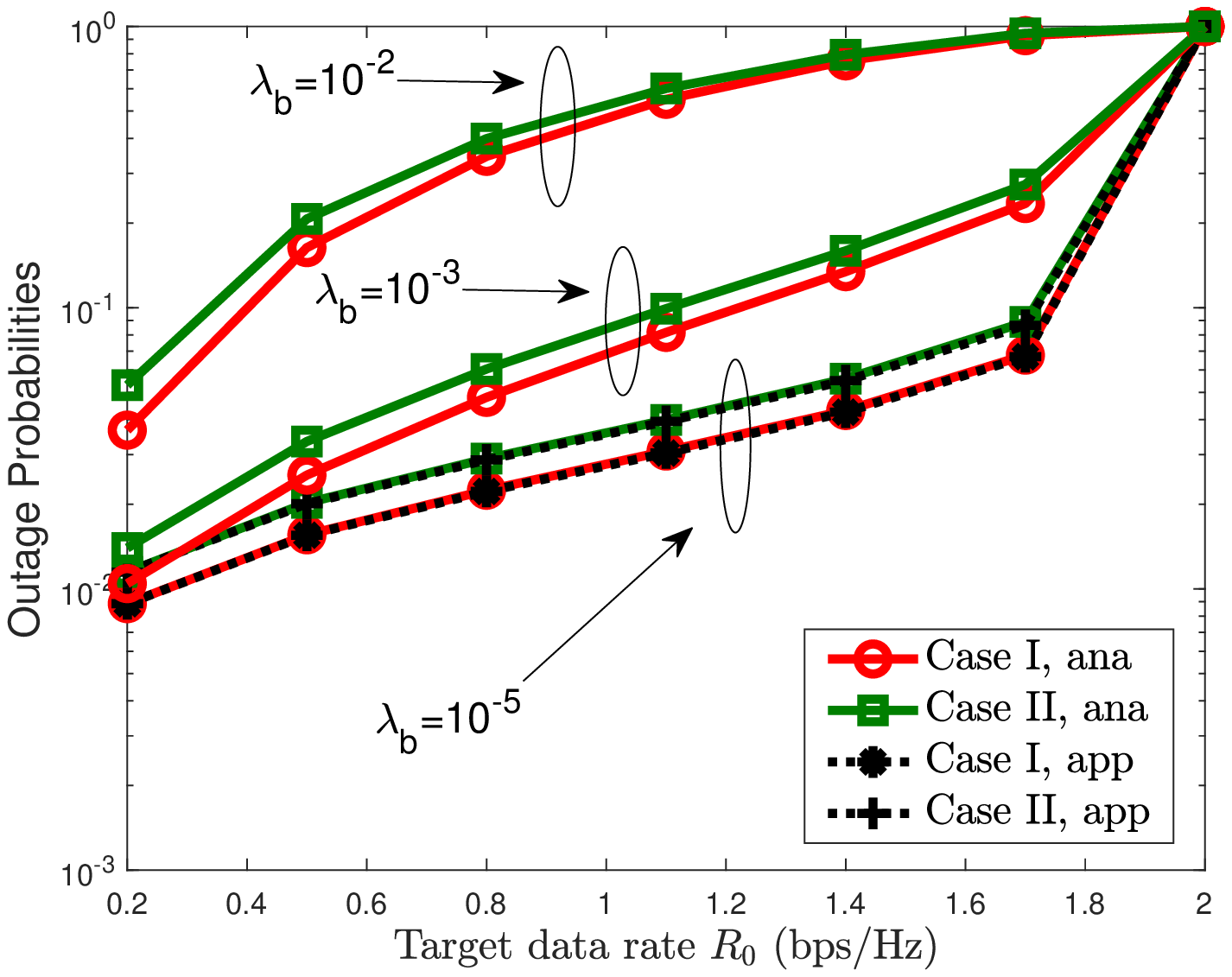}%
\label{impact_lambda2_CoMP}}
\hfil
\subfloat[NOMA user]{\includegraphics[width=3.2in]{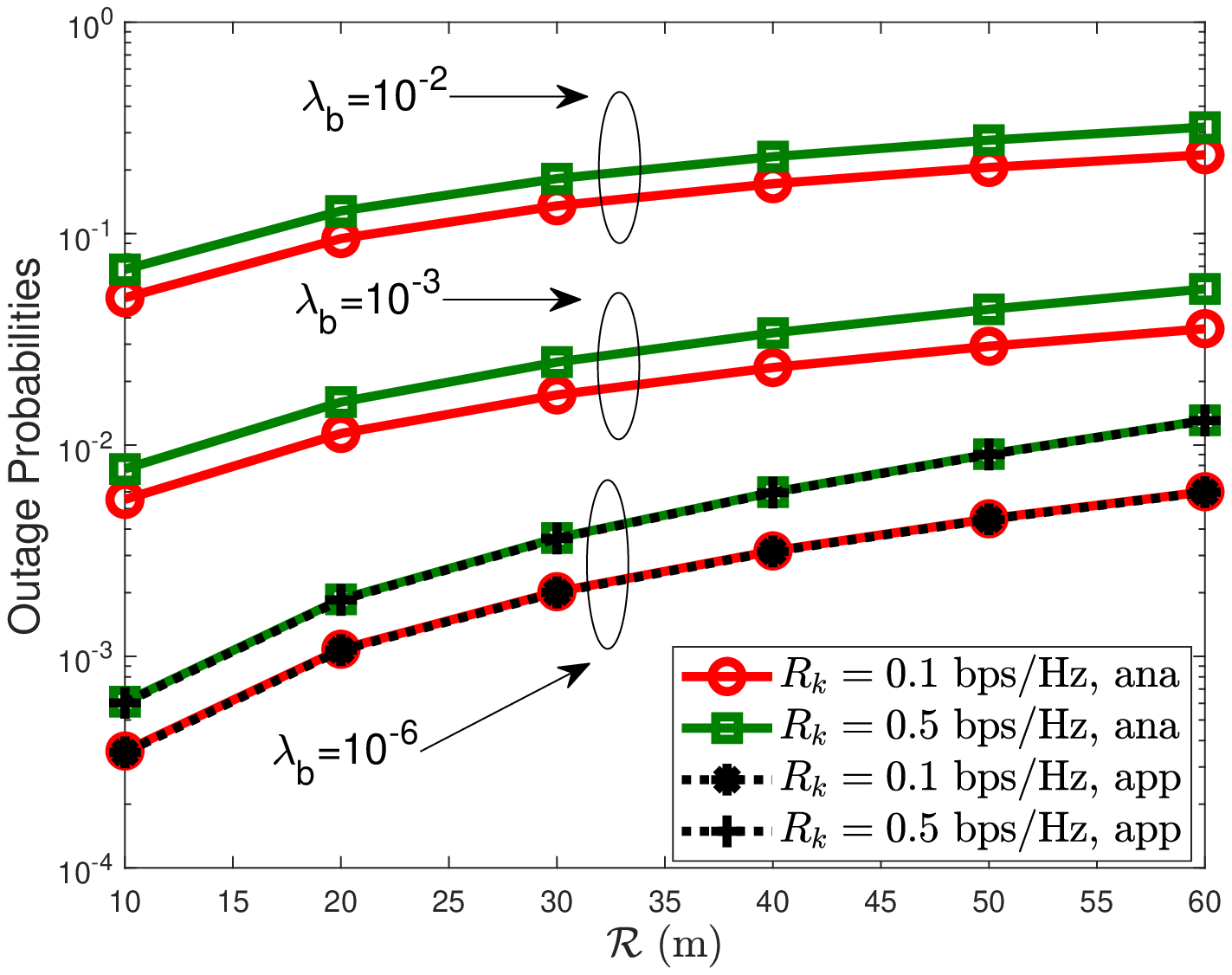}%
\label{impact_lambda2_NOMA}}
\caption{Impact of $\lambda_l$ and $\lambda_b$ on the performance when the product of $\lambda_l$ and
$\lambda_b$ is $2.5\times10^{-6}$. $\alpha_1=4$, $\beta=\frac{1}{4}$. (a) Case I: $d_1=d_2=100$ m; Case II: $d_1=100$ m, $d_2=150$ m. (b) $d_1=d_2=100$ m, $R_0=0.5$ bps/Hz.}
\label{impact_lambda2}
\end{figure*}

Figs. \ref{impact_lambda1} and \ref{impact_lambda2} show the impact of the intensity $\lambda_l$ of the representation space $\mathcal{C}$ and the intensity $\lambda_b$ of the nodes on each line on the  performance.  Fig. \ref{impact_lambda1} shows the outage probabilities versus $\lambda_b$ for different {\color{black}choices of $\lambda_l$}. It is noteworthy that $d_1=d_2=1/{2\lambda_b}$ is considered in Fig. \ref{impact_lambda1}, since $1/{\lambda_b}$ is the mean distance between two neighboring nodes on a line of the PLP. It is shown in Fig. \ref{impact_lambda1} that the outage probability achieved by the CoMP user decreases with $\lambda_b$, while that achieved by the NOMA user has the opposite trend. Fig. \ref{impact_lambda2} studies the case when the product of $\lambda_l$ and $\lambda_b$ is fixed. As shown in Fig. \ref{impact_lambda2}(a) and Fig. \ref{impact_lambda2}(b), when $\lambda_b$ is very small, the exact analytical results {\color{black}match} the curves of the approximated results. This verifies the {\color{black}results} obtained in Corollaries $1$ and $2$.

\section{Conclusions}
In order to improve the spectral efficiency, this paper has studied the application of  N-NOMA to vehicular communication networks. In the proposed scheme, CoMP users and NOMA users are served simultaneously in the same resource block. To take the topology of the vehicular network into consideration, a PLP driven Cox point process has been applied to model the {\color{black}locations} of the BSs and users along the roads. The Laplace transform of the interference has been derived, based on which the closed form expressions for the outage probabilities have been developed to provide system level performance. Asymptotic analytical results have also been provided for the case when the density of roads goes infinity and the density of BSs on each road goes zero. Extensive numerical results have been provided to show the performance achieved by the proposed scheme
in the vehicular network. It has been shown that the spectral efficiency achieved by the proposed N-NOMA scheme is much higher than that achieved by the conventional OMA scheme.

\appendices
\section{Proof for Lemma 1}
According to Assumption $1$, the interference observed at user $k$ at time slot
$t$ as shown in expression (\ref{I_k(t)}) can be written as follows:
\begin{align}\label{App_1}
I_k(t)=\sum\limits_{y_i\in\Omega_1}h_{i,k}z_i(t)+
       \sum\limits_{\{y_u,y_v\}\in\Omega_2}\left(h_{u,k}z_u(t)+h_{v,k}z_v(t)\right),
\end{align}
where $h_{i,k}$, $h_{u,k}$ and $h_{v,k}$ denote the channels from the interfering BSs to user $k$, whose specific forms {\color{black}are given by} either in expression (4) or expression (5) according to whether the interfering BSs locate on the typical line $l_0$;
$z_i(t)$, $z_u(t)$ and $z_v(t)$ are the transmitted signals, whose specific forms are different {\color{black}and} described in the following:
\begin{enumerate}
	\item $z_i(t)$ is the transmitted signal by
	a BS  which belongs to $\Omega_1$. According to Assumption $1$,
	$z_i(t)$ is independent for different $i$ and $t$. It is assumed that
	$\mathbb{E}\{|z_i(t)|^2\}=P$.
	\item $z_u(t)$ and $z_v(t)$ are the transmitted signals by a pair of BSs
    which belong to $\Omega_2$. According to Assumption $1$, the two
	cooperating BSs apply the proposed N-NOMA scheme, hence $z_u(t)$
	and $z_v(t)$ can be assumed to have the following forms:
	\begin{align}
      z_u(1)=\bar{z}_u(1)+\omega(1),\\\notag
      z_u(2)=\bar{z}_u(2)-\omega^*(2),\\\notag
      z_v(1)=\bar{z}_v(1)+\omega(2),\\\notag
      z_v(2)=\bar{z}_v(2)+\omega^*(1)
    \end{align}
    where $\bar{z}_p(t)$ ($p\in\{u,v\}$, $t\in\{1,2\}$) are independent signals intended for the NOMA users, and $\omega(j)$ ($j=1,2$) are the signals intended for the CoMP user.
    It is assumed that $\mathbb{E}\{|\bar{z}_p(t)|^2\}=\beta P$, and
    $\mathbb{E}\{|\omega(t)|^2\}=(1-\beta) P$.
\end{enumerate}
Then according to (19) and (\ref{App_1}).  The expectation of the power of
 $\tilde{I}_{k,p}$ can be calculated as
 \begin{align}
  \mathbb{E}\{|\tilde{I}_{k,p}|^2\}
    =\sum\limits_{y_i\in\Omega_1}|\eta_{i,p}|^2
    +\sum\limits_{\{y_u,y_v\}\in\Omega_2}|\eta_{u,v,p}|^2
 \end{align}
 where
 \begin{align}
  \eta_{i,p}=\theta_{k,p,1}h_{i,k}z_i(1)+\theta_{k,p,2}h_{i,k}^*z_i^*(2),
 \end{align}
 and
 \begin{align}
 \eta_{u,v,p}=&\theta_{k,p,1}(h_{u,k}\bar{z}_{u}(1)+h_{v,k}\bar{z}_{v}(1))
              +\theta_{k,p,2}(h_{u,k}^*\bar{z}_{u}^*(2)+h_{v,k}^*\bar{z}_{v}^*(2))\\\notag
              &+(\theta_{k,p,1}h_{u,k}+\theta_{k,p,2}h_{v,k}^*)\omega(1)
              +(\theta_{k,,p,1}h_{v,k}-\theta_{k,p,2}h_{u,k}^*)\omega(2).
 \end{align}
 Further, by noting that
 \begin{align}
  |\theta_{k,p,1}|^2+|\theta_{k,p,2}|^2=1,\\\notag
 \end{align}
 we have
 \begin{align}
 |\eta_{i,p}|^2=P|h_{i,k}|^2
 \end{align}
 and
 \begin{align}
  |\eta_{u,v,p}|=P(|h_{u,k}|^2+|h_{v,k}|^2).
 \end{align}
 Thus we have
  \begin{align}
  \mathbb{E}\{|\tilde{I}_{k,p}|^2\}
    &=\sum\limits_{y_i\in\Omega_1}P|h_{i,k}|^2
    +\sum\limits_{\{y_u,y_v\}\in\Omega_2}P(|h_{u,k}|^2+|h_{v,k}|^2)\\\notag
    &=P\left(\sum_{j\in\{L,R\}}\sum_{m=2,3,\cdots}\left|\tilde{h}_{j,m,k}\right|^2
   +\sum_{l_i\in\Phi_l}\sum_{x_{m}^{l_i}\in{\Psi_{l_i}^b}}\left|\hat{h}_{i,m,k}\right|^2\right),
 \end{align}
 where the last step follows by the fact that $\Psi_{l_0}^b\backslash \{x_{L,1}^{l_0},x_{R,1}^{l_0}\}
 \cup\underset{l_i\in\Phi_l}{\cup}\Psi_{l_i}^b=\Omega_1\cup\Omega_2$. {\color{black}The} proof for Lemma 1 is complete.

 \section{Proof for Lemma 2}
 The Laplace transform of $\zeta_{0}^{\text{intra}}$ can be calculated as follows:
 \begin{align}
 \mathcal{L}^{\text{intra}}(s,d_1,d_2)&=
 \mathbb{E}\left\{\exp\left(-s\sum_{j\in\{L,R\}}\sum_{m=2,3,\cdots}
                              \frac{|\tilde{g}_{j,m,0}|^2}{||x_{j,m}^{l_0}||^{\alpha_0}}\right)\right\}\\\notag
  &\overset{(a)}{=}\mathbb{E}\left\{\prod_{j\in\{L,R\}}\prod_{m=2,3,\cdots}
                 \frac{1}{1+\frac{s}{||x_{j,m}^{l_0}||^{\alpha_0}}}\right\}\\\notag
  &\overset{(b)}{=}\mathbb{E}\left\{\prod_{m=2,3,\cdots}
                 \frac{1}{1+\frac{s}{||x_{L,m}^{l_0}||^{\alpha_0}}}\right\}
     \cdot
     \mathbb{E}\left\{\prod_{m=2,3,\cdots}
                 \frac{1}{1+\frac{s}{||x_{R,m}^{l_0}||^{\alpha_0}}}\right\}\\\notag
  &\overset{(c)}{=}\exp\left(-\lambda_b\int_{d_1}^{\infty}1-\frac{1}{1+\frac{s}{r^{\alpha_0}}}\,dr\right)
     \cdot
     \exp\left(-\lambda_b\int_{d_2}^{\infty}1-\frac{1}{1+\frac{s}{r^{\alpha_0}}}\,dr\right)\\\notag
  &\overset{(d)}{=}\exp\bigg(-\lambda_b
  \bigg(
         \frac{sd_1^{1-\alpha_0}}{\alpha_0-1}{}_2F_1\left(1,1-\frac{1}{\alpha_0};2-\frac{1}{\alpha_0};-\frac{s}{d_1^{\alpha_0}}\right)\\\notag
         &\quad\quad\quad\quad+
         \frac{sd_2^{1-\alpha_0}}{\alpha_0-1}{}_2F_1\left(1,1-\frac{1}{\alpha_0};2-\frac{1}{\alpha_0};-\frac{s}{d_2^{\alpha_0}}\right)
  \bigg)\bigg),
 \end{align}
 where step (a) follows by the fact that the small scale fading gains $|\tilde{g}_{j,m,0}|^2$
 are i.i.d exponential random variables with parameter $1$, step (b) follows by the fact
 the nodes on the left hand side of the origin and the nodes on the right hand side of the origin are independently {\color{black}located}, step (c) follows by applying the probability generating functional (PGFL) of the 1D HPPPs on the left and right hand side of the origin,
 respectively, and step (d) follows by applying the Gauss hyper-geometric function.
 \section{Proof for Lemma 3}
 The Laplace transform of $\zeta_{0}^{\text{inter}}$ can be calculated as follows:
 \begin{align}
  \mathcal{L}^{\text{inter}}(s)&=
  \mathbb{E}\left\{\exp\left(-s\sum_{l_i\in\Phi_l}\sum_{x_{m}^{l_i}\in{\Psi_{l_i}^b}}
                     \frac{|\hat{g}_{i,m,0}|^2}{||x_m^{l_i}||^{\alpha_1}}\right)\right\}\\\notag
  &\overset{(a)}{=}\mathbb{E}\left\{\prod_{l_i\in\Phi_l}\prod_{x_{m}^{l_i}\in{\Psi_{l_i}^b}}
                \frac{1}{1+\frac{s}{||x_m^{l_i}||^{\alpha_1}}}\right\}\\\notag
  &\overset{(b)}{=}\mathbb{E}\left\{\prod_{l_i\in\Phi_l}
         \mathbb{E}\left\{\prod_{x_{m}^{l_i}\in{\Psi_{l_i}^b}}
                \frac{1}{1+\frac{s}{(\rho_i^2+u_{i,m}^2)^{\alpha_1/2}}}\bigg|l_i\right\}\right\}\\\notag
  &\overset{(c)}{=}\mathbb{E}\left\{\prod_{l_i\in\Phi_l}
                 \exp\left(-2\lambda_b\int_0^{\infty}\frac{s}{s+(\rho_i^2+u^2)^{\alpha_1/2}}\,du\right)\right\}\\\notag
  &\overset{(d)}{=}\exp\left(-2\pi\lambda_l\int_0^{\infty}1-
                 \exp\left(-2\lambda_b\int_0^{\infty}\frac{s}{s+(x^2+u^2)^{\alpha_1/2}}\,du\right)\,dx\right)
 \end{align}
where step (a) follows by the fact that the small scale fading gains
$|\hat{g}_{i,m,0}|^2$
 are i.i.d exponential random variables with parameter $1$,
$\rho_i$ in (b) is the perpendicular distance from the origin to
line $l_i$, $u_{i,m}$ is the  distance from the $m$-th node
on line $l_i$ to the projection of the origin onto line $l_i$,
step (c) follows by applying the PGFL of the 1D HPPP on each line $l_i$,
and step (d) follows by applying the PGFL of the 2D HPPP $\Xi$ on the representation  space $\mathcal{C}$.

\section{Proof for Theorem 1}
Given $d_1$ and $d_2$, the outage probability achieved by the CoMP user
can be written as follows:
\begin{align}
P_0^{out}(d_1,d_2)=P\left(C_0^2<\frac{\epsilon_0}{1-\beta-\epsilon_0\beta}(\zeta_0+\frac{1}{\rho})\right).
\end{align}
Note that, $C_0^2=|\tilde{h}_{L,1,0}|^2+|\tilde{h}_{R,1,0}|^2$, and $|\tilde{h}_{L,1,0}|^2$ and $|\tilde{h}_{R,1,0}|^2$ are exponentially distributed with parameters $d_1^{\alpha_0}$ and $d_2^{\alpha_0}$, respectively. {\color{black}Therefore,} the CDF of
$C_0^2$ can be easily obtained as given by the following two cases
\begin{itemize}
	\item when $d_1\neq d_2$,
	  \begin{align}
	     F_{C_0^2}(x)=1-\frac{d_2^{\alpha_0}}{d_2^{\alpha_0}-d_1^{\alpha_0}}
	                     e^{-d_1^{\alpha_0}x}
	                   +\frac{d_1^{\alpha_0}}{d_2^{\alpha_0}-d_1^{\alpha_0}}
	                     e^{-d_1^{\alpha_0}x},
	   \end{align}
	\item when $d_1=d_2$
	 \begin{align}
	    F_{C_0^2}=1-e^{-d_1^{\alpha_0}x}-d_1^{\alpha_0}xe^{-d_1^{\alpha_0}x}.
	 \end{align}
\end{itemize}

Then the outage probability can be calculated as follows:
\begin{itemize}
	\item when $d_1\neq d_2$,
\begin{align}
P_0^{out}(d_1,d_2)=&1-\frac{d_2^{\alpha_0}}{d_2^{\alpha_0}-d_1^{\alpha_0}}
             e^{-\frac{\mu(\epsilon_0,d_1)}{\rho}}
             \mathbb{E}_{\zeta_0}\left\{e^{-\mu(\epsilon_0,d_1)\zeta_0}\right\}\\\notag
          &+\frac{d_1^{\alpha_0}}{d_2^{\alpha_0}-d_1^{\alpha_0}}
             e^{-\frac{\mu(\epsilon_0,d_2)}{\rho}}
             \mathbb{E}_{\zeta_0}\left\{e^{-\mu(\epsilon_0,d_2)\zeta_0}\right\}.
\end{align}
By applying the Laplace transform of $\zeta_0$, the expression in (\ref{P1_ne}) can be obtained.
	\item when $d_1=d_2$
	 \begin{align}
	   P_0^{out}(d_1,d_2)=&1-e^{-\frac{-\mu(\epsilon_0,d_1)}{\rho}}\left(1+\frac{\mu(\epsilon_0,d_1)}{\rho}\right)
	                      \mathbb{E}_{\zeta_0}\left\{e^{-\mu(\epsilon_0,d_1)\zeta_0}\right\}\\\notag
	    &\quad-e^{-\frac{\mu(\epsilon_0,d_1)}{\rho}}\mu(\epsilon_0,d_1)\mathbb{E}_{\zeta_0}\left\{\zeta_0e^{-\mu(\epsilon_0,d_1)\zeta_0}\right\}.
	 \end{align}

	 By applying the Laplace transform of $\zeta_0$ and the following relationship
	 \begin{align}
	  \mathbb{E}_{\zeta_0}\left\{\zeta_0e^{-\mu(\epsilon_0,d_1)\zeta_0}\right\}=
	    -\frac{\partial \mathcal{L}(\mu(\epsilon_0,d_1),d_1,d_1)}{\partial \mu(\epsilon_0,d_1)},
	 \end{align}
	 the expression in (\ref{P1_e}) can be obtained and the proof for Theorem 1 is complete.
\end{itemize}

\section{Proof for Theorem 2}
It is worth pointing out that this section only {\color{black}includes} the proof for user $1$, since the results for user $2$ can be proved by following the same procedure as user $1$.

For notational simplicity, we denote the distance from user $1$ to its associating BS locating at $x_{L,1}^{l_0}$ by $r_1=||x_{L,1}^{l_0}-U_1||$, and the distance  from user $1$ to the interfering BS locating at $x_{R,1}^{l_0}$ by $r_2=||x_{R,1}^{l_0}-U_1||$.

One key step to calculate the outage probability is to obtain the Laplace transform for the interference $\zeta_1$. Note that, the Laplace transform for $\zeta_1$ is dependent on the location of user $1$, and can be expressed into two cases as highlighted in the following Lemma.
\begin{Lemma}
Given $d_1$, $d_2$ and $r_1$, the Laplace transform for $\zeta_1$ can be expressed as follows:
\begin{itemize}
	\item when $U_1$ locates at the right hand side of $x_{L,1}^{l_0}$,
	 \begin{align}\label{Lap_NOMA11}
	  \mathbb{E}\{e^{-s\zeta_1}\}&=\mathcal{L}_{1,1}(s,d_1,d_2,r_1)\\\notag
	                            &=\mathcal{L}_0^{\text{intra}}(s,r_1,d_1+d_2-r_1)
	                              \mathcal{L}^{\text{inter}}(s),
	 \end{align}
	\item when $U_1$ locates at the left hand side of $x_{L,1}^{l_0}$,
     \begin{align}\label{Lap_NOMA12}
	  \mathbb{E}\{e^{-s\zeta_1}\}&=\mathcal{L}_{1,2}(s,d_1,d_2,r_1)\\\notag
	                            &=\exp\left(-2\lambda_b\int_0^{r_1}\frac{s}{s+x^
	                            {\alpha_0}}\,dx\right)
	                            \mathcal{L}_0^{\text{intra}}(s,r_1,d_1+d_2+r_1)
	                              \mathcal{L}^{\text{inter}}(s).
	 \end{align}
\end{itemize}
\end{Lemma}
Note that, the proof for the above lemma is similar to the CoMP user case and is omitted in this paper.
\begin{figure*}[!t]
\vspace{-3em}
\centering
\subfloat[Case I]{\includegraphics[width=2in]{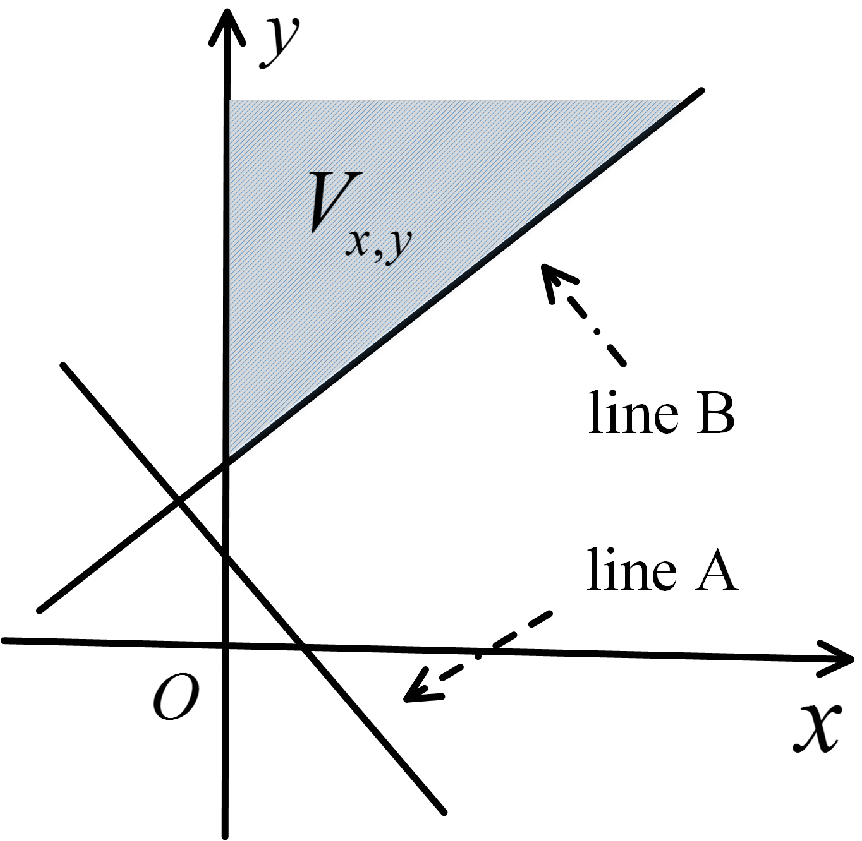}}
\hfil
\subfloat[Case II]{\includegraphics[width=2in]{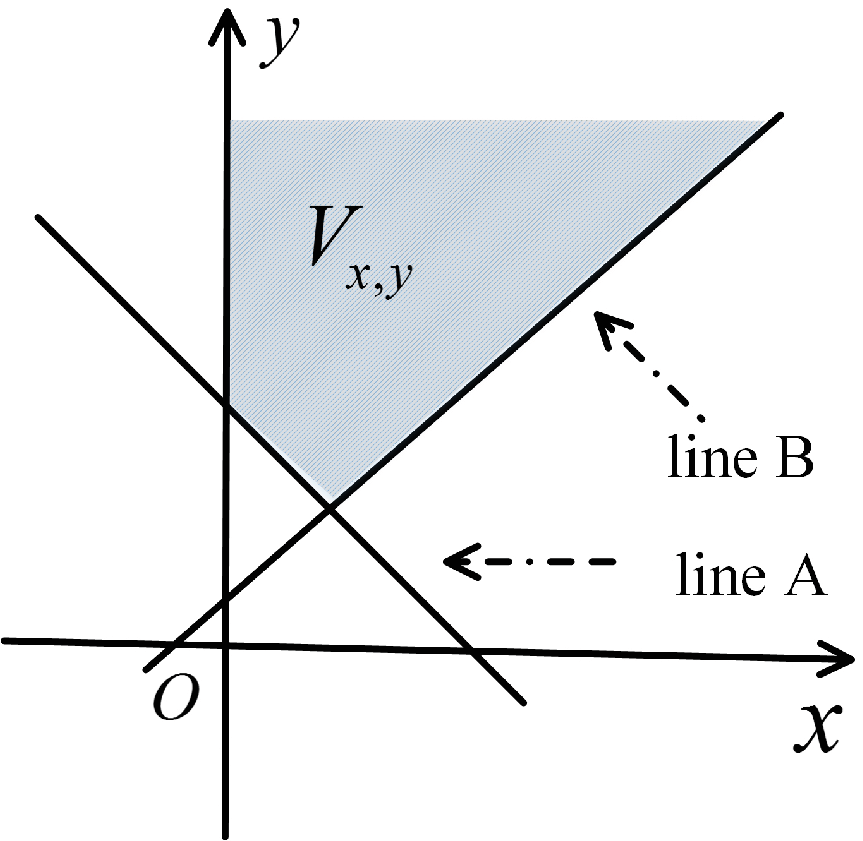}}%
\caption{Illustration of the integration region. Line A: $y=-x+\frac{\epsilon_0(\zeta_1+1/\rho)}{1-\beta-\epsilon_0\beta}$, line B: $y=\epsilon_1x+\frac{\epsilon_1(\zeta_1+1/\rho)}{\beta}$}
\label{proof_theorem2}
\end{figure*}

Given $d_1$ and $d_2$, the outage probability achieved by user $1$ can be rewritten  as
\begin{align}
P_1^{out}(d_1,d_2)=1-P\bigg(&|\tilde{h}_{L,1,1}|^2>-|\tilde{h}_{R,1,1}|^2
                            +\frac{\epsilon_0(\zeta_1+1/\rho)}{1-\beta-\epsilon_0\beta},\\\notag
          &|\tilde{h}_{L,1,1}|^2>\epsilon_1|\tilde{h}_{R,1,1}|^2+
           \frac{\epsilon_1(\zeta_1+1/\rho)}{\beta}\bigg).
\end{align}
Then $P_1^{out}(d_1,d_2)$ can be further calculated as follows:
\begin{align}\label{P1_int}
P_1^{out}(d_1,d_2)=1-\mathbb{E}_{U_1,\zeta_1}\left\{
         \underset{Q_{\text{int}}}{\underbrace{\underset{(x,y)\in V_{x,y}}{\iint}f_{|\tilde{h}_{R,1,1}|^2}(x)f_{|\tilde{h}_{L,1,1}|^2}(y)\,dxdy}}\right\},
\end{align}
where $f_{|\tilde{h}_{R,1,1}|^2}(x)$ and $f_{|\tilde{h}_{L,1,1}|^2}(y)$ are the {\color{black}pdfs} of $|\tilde{h}_{L,1,1}|^2$ and $|\tilde{h}_{R,1,1}|^2$, respectively, which are given by
\begin{align}
 f_{|\tilde{h}_{R,1,1}|^2}(x)=r_2^{\alpha_0}e^{-r_2^{\alpha_0}x},\\
 f_{|\tilde{h}_{L,1,1}|^2}(y)=r_1^{\alpha_0}e^{-r_1^{\alpha_0}y},
\end{align}
and $V_{x,y}$ is the integral region, which can be divided into  two cases determined by the relationship among $\beta$, $\epsilon_0$ and $\epsilon_1$ as follows:
\begin{itemize}
	\item Case I: when $\epsilon_1\geq\frac{\epsilon_0\beta}{1-\beta-\epsilon_0\beta}$, as shown in Fig. \ref{proof_theorem2}(a),
	\begin{align}\label{region1}
	 V_{x,y}=\left\{(x,y)|x\in\mathbb{R}^+, \epsilon_1x+\frac{\epsilon_1(\zeta_1+1/\rho)}{\beta}<y<+\infty\right\},
	\end{align}
	where $\mathbb{R}^+$ is the positive real number set.
	\item Case II: when $\epsilon_1 < \frac{\epsilon_0\beta}{1-\beta-\epsilon_0\beta}$, as shown in Fig. \ref{proof_theorem2}(b),
	 \begin{align}
	 V_{x,y}=&\left\{(x,y)|0<x<\eta(\zeta_1+1/\rho), -x+\frac{\epsilon_0(\zeta_1+1/\rho)}{1-\beta-\epsilon_0\beta}<y<+\infty\right\} \cup\\\notag
	         &\left\{(x,y)|\eta(\zeta_1+1/\rho)<x<+\infty, \epsilon_1x+\frac{\epsilon_1(\zeta_1+1/\rho)}{\beta}<y<+\infty\right\},
	\end{align}
	where $\eta=\frac{\epsilon_0\beta-\epsilon_1(1-\beta-\epsilon_0\beta)}
	                {\beta(1-\beta-\epsilon_0\beta)(1+\epsilon_1)}$. Note that
	 $\eta(\zeta_1+1/\rho)$ is the x-coordinate of the intersection of the two lines A: $y=-x+\frac{\epsilon_0(\zeta_1+1/\rho)}{1-\beta-\epsilon_0\beta}$
	 and B: $y=\epsilon_1x+\frac{\epsilon_1(\zeta_1+1/\rho)}{\beta}$.
\end{itemize}

In the following, we will concentrate on the calculation for Case I. Based on
(\ref{region1}), the integration in (\ref{P1_int}) can be calculated as
\begin{align}
Q_{\text{int}}=\frac{r_2^{\alpha_0}}{\epsilon_1r_1^{\alpha_0}+r_2^{\alpha_0}}
                \exp\left(-\frac{r_1^{\alpha_0}\epsilon_1}{\beta\rho}\right)
                \exp\left(-\frac{r_1^{\alpha_0}\epsilon_1\zeta_1}{\beta}\right).
\end{align}

Then the outage probability can be expressed as
\begin{align}\label{EE}
P_1^{out}(d_1,d_2)=1-\mathbb{E}_{U_1}\left\{\underset{E}{\underbrace{\mathbb{E}_{\zeta_1}\left\{
                      \frac{r_2^{\alpha_0}}{\epsilon_1r_1^{\alpha_0}+r_2^{\alpha_0}}
                \exp\left(-\frac{r_1^{\alpha_0}\epsilon_1}{\beta\rho}\right)
                \exp\left(-\frac{r_1^{\alpha_0}\epsilon_1\zeta_1}{\beta}\right)\bigg|U_1\right\}}}\right\}.
\end{align}
The calculation for the conditional expectation can be divided into the following two cases:
\begin{itemize}
	\item when $U_1$ locates at the right hand side of $x_{L,1}^{l_0}$, we have
	   $r_2=d_1+d_2-r_1$. By applying the Laplace transform of $\zeta_1$ as expressed in (\ref{Lap_NOMA11}), $E$ can be expressed as
\begin{align}\label{E1}
E=\frac{(d_1+d_2-r_1)^{\alpha_0}}{(d_1+d_2-r_1)^{\alpha_0}+\epsilon_1r_1^{\alpha_0}}
   \exp\left(-\frac{r_1^{\alpha_0}\epsilon_1}{\beta\rho}\right)\mathcal{L}_{1,1}(\frac{r_1^{\alpha_0}\epsilon_1}{\beta},d_1,d_2,r_1)
\end{align}
    \item when $U_1$ locates at the left hand side of $x_{L,1}^{l_0}$,  we have
	   $r_2=d_1+d_2+r_1$. By applying the Laplace transform of $\zeta_1$ as expressed in (\ref{Lap_NOMA12}), $E$ can be expressed as
\begin{align}\label{E2}
E=\frac{(d_1+d_2+r_1)^{\alpha_0}}{(d_1+d_2+r_1)^{\alpha_0}+\epsilon_1r_1^{\alpha_0}}
   \exp\left(-\frac{r_1^{\alpha_0}\epsilon_1}{\beta\rho}\right)\mathcal{L}_{1,2}(\frac{r_1^{\alpha_0}\epsilon_1}{\beta},d_1,d_2,r_1)
\end{align}
By taking (\ref{E1}) and (\ref{E2}) into (\ref{EE}) and applying the Gauss-Chebyshev quadrature, the proof for case I is complete.
\end{itemize}
\section{Proof for Corollaries $1$ and $2$}
{\color{black}In this section}, only the proof for the case when $d_1\neq d_2$ of the CoMP user is provided, since {\color{black}the proofs for} other cases can be {\color{black}obtained} by following the similar procedure.

When $\lambda_l\to \infty$ and $\lambda_l\lambda_b=\lambda$, the outage probability achieved by the CoMP user can be expressed as
\begin{align}
 \lim_{\lambda_l\to\infty}P_0^{out}(d_1,d_2)=&1-\frac{d_2^{\alpha_0}}{d_2^{\alpha_0}-d_1^{\alpha_0}}
             e^{-\frac{\mu(\epsilon_0,d_1)}{\rho}}
            \lim_{\lambda_l\to\infty} \mathcal{L}\left(\mu(\epsilon_0,d_1),d_1,d_2\right)\\\notag
          &+\frac{d_1^{\alpha_0}}{d_2^{\alpha_0}-d_1^{\alpha_0}}
             e^{-\frac{\mu(\epsilon_0,d_2)}{\rho}}
            \lim_{\lambda_l\to\infty} \mathcal{L}\left(\mu(\epsilon_0,d_2),d_1,d_2\right),
\end{align}
Thus the key is to calculate the limit of the Laplace transform. It can be easily obtained from  expression (\ref{Lap_intra_0}) that
\begin{align}
 \lim_{\lambda_l\to\infty}\mathcal{L}^{\text{intra}}_0(s,d_1,d_2)=1.
\end{align}
The limit of the interline interference can be evaluated as follows:
\begin{align}
\lim_{\lambda_l\to \infty}\mathcal{L}^{\text{inter}}(s)
&=\exp\left(\lim_{\lambda_l\to \infty}-2\pi\lambda_l\int_0^{\infty}1-
                 \exp\left(-\frac{2\lambda}{\lambda_l}\underset{Q(x)}{\underbrace{\int_0^{\infty}\frac{s}{s+(x^2+u^2)^{\alpha_1/2}}\,du}}\right)\,dx\right)\\\notag
&=\exp\left(-4\pi\lambda\int_{0}^{\infty}Q(x)\,dx+
            \underset{\Theta}{\underbrace{\lim_{\lambda_l\to \infty}2\pi\int_{0}^{\infty}\sum_{n=2}^{\infty}\frac{(-2\lambda)^nQ^n(x)}{n!}\frac{1}{\lambda_l^{n-1}}\,dx}}\right),
\end{align}
where the last step follows by Taylor expansion. Further, the limit term $\Theta$ can be calculated as follows:
\begin{align}
\Theta
&\overset{(a)}{=}2\pi\int_0^{\infty}\lim_{\lambda_l\to \infty}\sum_{n=2}^{\infty}\frac{(-2\lambda)^nQ^n(x)}{n!}\frac{1}{\lambda_l^{n-1}}\,dx\\\notag
&\overset{(b)}{=}2\pi\int_0^{\infty}\sum_{n=2}^{\infty}\lim_{\lambda_l\to \infty}\frac{(-2\lambda)^nQ^n(x)}{n!}\frac{1}{\lambda_l^{n-1}}\,dx\\\notag
&=0,
\end{align}
where step (a) follows by applying the dominated convergence theorem \cite{Royden1988real}, step (b) follows by applying the
Tannery's theorem \cite{loya2017amazing}.
Therefore, we have
\begin{align}
\lim_{\lambda_l\to \infty}\mathcal{L}^{\text{inter}}(s)
&=\exp\left(-4\pi\lambda\int_{0}^{\infty}Q(x)\,dx\right)
=\bar{\mathcal{L}}(s).
\end{align}
Finally, it is obtained that
\begin{align}
\lim_{\lambda_l\to\infty} \mathcal{L}\left(s,d_1,d_2\right)
&=\lim_{\lambda_l\to\infty}\mathcal{L}^{\text{intra}}_0(s,d_1,d_2) \lim_{\lambda_l\to \infty}\mathcal{L}^{\text{inter}}(s)\\\notag
&=\bar{\mathcal{L}}(s),
\end{align}
and the proof is complete.
\bibliographystyle{IEEEtran}
\bibliography{IEEEabrv,ref}

\begin{thebibliography}{10}
\providecommand{\url}[1]{#1}
\csname url@samestyle\endcsname
\providecommand{\newblock}{\relax}
\providecommand{\bibinfo}[2]{#2}
\providecommand{\BIBentrySTDinterwordspacing}{\spaceskip=0pt\relax}
\providecommand{\BIBentryALTinterwordstretchfactor}{4}
\providecommand{\BIBentryALTinterwordspacing}{\spaceskip=\fontdimen2\font plus
\BIBentryALTinterwordstretchfactor\fontdimen3\font minus
  \fontdimen4\font\relax}
\providecommand{\BIBforeignlanguage}[2]{{%
\expandafter\ifx\csname l@#1\endcsname\relax
\typeout{** WARNING: IEEEtran.bst: No hyphenation pattern has been}%
\typeout{** loaded for the language `#1'. Using the pattern for}%
\typeout{** the default language instead.}%
\else
\language=\csname l@#1\endcsname
\fi
#2}}
\providecommand{\BIBdecl}{\relax}
\BIBdecl

\bibitem{saito2013non}
Y.~Saito, Y.~Kishiyama, A.~Benjebbour, T.~Nakamura, A.~Li, and K.~Higuchi,
  ``{Non-orthogonal multiple access (NOMA) for cellular future radio access},''
  in \emph{Proc. {IEEE} Veh. Technol. Conf.}, Dresden, Germany, Jun. 2013, pp.
  1--5.

\bibitem{ding2017NOMAsurvey}
Z.~Ding, X.~Lai, G.~K. Karagiannidis, R.~Schober, J.~Yuan, and V.~Bhargava,
  ``{A survey on non-orthogonal multiple access for 5G networks: research
  challenges and future trends},'' \emph{{IEEE} J. Sel. Areas Commun.},
  vol.~35, no.~10, pp. 2181--2195, Oct. 2017.

\bibitem{islam2017noma5G}
S.~Islam, M.~Zeng, and O.~A. Dobre, ``{NOMA in 5G systems: Exciting
  possibilities for enhancing spectral efficiency},'' \emph{IEEE 5G Tech.
  Focus}, vol.~1, no.~2, pp. 1--6, Jun. 2017.

\bibitem{ding2017application}
Z.~Ding, Y.~Liu, J.~Choi, Q.~Sun, M.~Elkashlan, I.~Chih-Lin, and H.~V. Poor,
  ``{Application of non-orthogonal multiple access in LTE and 5G networks},''
  \emph{{IEEE} Commun. Mag.}, vol.~55, no.~2, pp. 185--191, Feb. 2017.

\bibitem{ding2016general}
Z.~Ding, R.~Schober, and H.~V. Poor, ``{A general MIMO framework for NOMA
  downlink and uplink transmission based on signal alignment},'' \emph{{IEEE}
  Trans. Wireless Commun.}, vol.~15, no.~6, pp. 4438--4454, Jun. 2016.

\bibitem{choi2015minimum}
J.~Choi, ``{Minimum power multicast beamforming with superposition coding for
  multiresolution broadcast and application to NOMA systems},'' \emph{{IEEE}
  Trans. Commun.}, vol.~63, no.~3, pp. 791--800, Mar. 2015.

\bibitem{octavia2017MIMONOMA}
M.~Zeng, A.~Yadav, O.~A. Dobre, G.~I. Tsiropoulos, and H.~V. Poor, ``{On the
  sum rate of MIMO-NOMA and MIMO-OMA systems},'' \emph{{IEEE} Wireless Commun.
  Lett.}, vol.~6, no.~4, pp. 534--537, Aug. 2017.

\bibitem{ding2017randommmwave}
Z.~Ding, P.~Fan, and H.~V. Poor, ``{Random beamforming in millimeter-wave NOMA
  networks},'' \emph{IEEE Access}, vol.~5, pp. 7667--7681, Feb. 2017.

\bibitem{yzhoummwvave2018}
Y.~{Zhou}, V.~W.~S. {Wong}, and R.~{Schober}, ``{Coverage and rate analysis of
  millimeter wave NOMA networks with beam misalignment},'' \emph{IEEE Trans.
  Wireless Commun.}, vol.~17, no.~12, pp. 8211--8227, Dec. 2018.

\bibitem{sysmmwave2018}
Y.~{Sun}, Z.~{Ding}, and X.~{Dai}, ``{On the performance of downlink NOMA in
  multi-cell mmWave networks},'' \emph{IEEE Commun. Lett.}, vol.~22, no.~11,
  pp. 2366--2369, Nov. 2018.

\bibitem{Ding2019MEC}
Z.~{Ding}, P.~{Fan}, and H.~V. {Poor}, ``{Impact of non-orthogonal multiple
  access on the offloading of mobile edge computing},'' \emph{{IEEE} Trans.
  Commun.}, vol.~67, no.~1, pp. 375--390, Jan. 2019.

\bibitem{choi2014non}
J.~Choi, ``{Non-orthogonal multiple access in downlink coordinated two-point
  systems},'' \emph{{IEEE} Commun. Lett.}, vol.~18, no.~2, pp. 313--316, Feb.
  2014.

\bibitem{ali2018coordinated}
M.~S. Ali, E.~Hossain, and D.~I. Kim, ``{Coordinated multipoint transmission in
  downlink multi-cell NOMA systems: Models and spectral efficiency
  performance},'' \emph{IEEE Wireless Commun.}, vol.~25, no.~2, pp. 24--31,
  Apr. 2018.

\bibitem{tian2016performance}
Y.~Tian, A.~R. Nix, and M.~Beach, ``{On the Performance of Opportunistic NOMA
  in Downlink CoMP Networks},'' \emph{{IEEE} Commun. Lett.}, vol.~20, no.~5,
  pp. 998--1001, May 2016.

\bibitem{venkatesan2007network}
S.~Venkatesan, A.~Lozano, and R.~Valenzuela, ``{Network MIMO: Overcoming
  Intercell Interference in Indoor Wireless Systems},'' in \emph{Proc. IEEE
  ACSSC¡¯07}, Pacific Grove, CA, USA, Nov. 2007, pp. 83--87.

\bibitem{huang2009increasing}
H.~Huang, M.~Trivellato, A.~Hottinen, M.~Shafi, P.~J. Smith, and R.~Valenzuela,
  ``{Increasing downlink cellular throughput with limited network MIMO
  coordination},'' \emph{{IEEE} Trans. Wireless Commun.}, vol.~8, no.~6, pp.
  2983--2989, June 2009.

\bibitem{jungnickel2009coordinated}
V.~Jungnickel, L.~Thiele, T.~Wirth, T.~Haustein, S.~Schiffermuller, A.~Forck,
  S.~Wahls, S.~Jaeckel, S.~Schubert, H.~Gabler \emph{et~al.}, ``{Coordinated
  multipoint trials in the downlink},'' in \emph{Proc. {IEEE} Globecom.
  Workshops}, Honolulu, HI, USA, Dec. 2009, pp. 1--7.

\bibitem{sys2017nnomafeasibility}
Y.~Sun, Z.~Ding, X.~Dai, and G.~K. Karagiannidis, ``{A feasibility study on
  network NOMA},'' \emph{{IEEE} Trans. Commun.}, vol.~66, no.~9, pp.
  4303--4317, Sep. 2018.

\bibitem{ali2018downlink}
M.~S. Ali, E.~Hossain, A.~Al-Dweik, and D.~I. Kim, ``{Downlink power allocation
  for CoMP-NOMA in multi-cell networks},'' \emph{IEEE Trans. Commun.}, vol.~66,
  no.~9, pp. 3982--3998, Sep. 2018.

\bibitem{sys2019PCP}
Y.~Sun, Z.~Ding, X.~Dai, and O.~A. {Dobre}, ``{On the performance of network
  NOMA in uplink CoMP systems: a stochastic geometry approach},'' \emph{{IEEE}
  Trans. Commun.}, vol.~67, no.~7, pp. 5084--5098, Jul. 2019.

\bibitem{sahin2018virtual}
T.~Sahin, M.~Klugel, C.~Zhou, and W.~Kellerer, ``{Virtual cells for 5G V2X
  communications},'' \emph{IEEE Commun. Stands Mag}, vol.~2, no.~1, pp. 22--28,
  Mar. 2018.

\bibitem{boban2018connected}
M.~Boban, A.~Kousaridas, K.~Manolakis, J.~Eichinger, and W.~Xu, ``{Connected
  roads of the future: Use cases, requirements, and design considerations for
  vehicle-to-everything communications},'' \emph{IEEE Veh. Technol. Mag.},
  vol.~13, no.~3, pp. 110--123, Jul. 2018.

\bibitem{chen2017vehicle}
S.~Chen, J.~Hu, Y.~Shi, Y.~Peng, J.~Fang, R.~Zhao, and L.~Zhao,
  ``{Vehicle-to-everything (V2X) services supported by LTE-based systems and
  5G},'' \emph{IEEE Commun. Stands. Mag.}, vol.~1, no.~2, pp. 70--76, Jun.
  2017.

\bibitem{GPP2016V2X}
\emph{{Study on LTE-based V2X services}}, {3GPP TR 36.885}, Jul. 2016.

\bibitem{chen2017performance}
Y.~Chen, L.~Wang, Y.~Ai, B.~Jiao, and L.~Hanzo, ``{Performance analysis of
  NOMA-SM in vehicle-to-vehicle massive MIMO channels},'' \emph{{IEEE} J.
  Select. Areas Commun.}, vol.~35, no.~12, pp. 2653--2666, Dec. 2017.

\bibitem{di2017non}
B.~Di, L.~Song, Y.~Li, and G.~Y. Li, ``{Non-orthogonal multiple access for
  high-reliable and low-latency V2X communications in 5G systems},''
  \emph{{IEEE} J. Select. Areas Commun.}, vol.~35, no.~10, pp. 2383--2397, Jul.
  2017.

\bibitem{zhang2019performance}
D.~Zhang, Y.~Liu, L.~Dai, A.~K. Bashir, A.~Nallanathan, and B.~Shim,
  ``{Performance analysis of FD-NOMA-based decentralized V2X systems},''
  \emph{IEEE Trans. Commun.}, vol.~67, no.~7, pp. 5024--5036, Jul. 2019.

\bibitem{choi2018poisson}
C.-S. Choi and F.~Baccelli, ``{Poisson cox point processes for vehicular
  networks},'' \emph{{IEEE} Trans. Veh. Technol.}, vol.~67, no.~10, pp.
  10\,160--10\,165, Jul. 2018.

\bibitem{choi2018analytical}
------, ``{An analytical framework for coverage in cellular networks leveraging
  vehicles},'' \emph{IEEE Trans. Commun.}, vol.~66, no.~10, pp. 4950--4964, May
  2018.

\bibitem{chetlur2018coverage}
V.~V. Chetlur and H.~S. Dhillon, ``{Coverage analysis of a vehicular network
  modeled as Cox process driven by Poisson line process},'' \emph{IEEE Trans.
  Wireless Commun.}, vol.~17, no.~7, pp. 4401--4416, Apr. 2018.

\bibitem{haenggi2012stochastic}
M.~Haenggi, \emph{{Stochastic geometry for wireless networks}}.\hskip 1em plus
  0.5em minus 0.4em\relax Cambridge, U.K.: Cambridge Univ. Press, 2012.

\bibitem{haenggi2015meta}
------, ``{The meta distribution of the SIR in Poisson bipolar and cellular
  networks},'' \emph{IEEE Trans. Wireless Commun.}, vol.~15, no.~4, pp.
  2577--2589, Dec. 2015.

\bibitem{Royden1988real}
H.~L. Royden, \emph{{Real analysis}}.\hskip 1em plus 0.5em minus 0.4em\relax
  New York: Macmillan, 1988.

\bibitem{loya2017amazing}
P.~Loya, \emph{{Amazing and aesthetic aspects of analysis}}.\hskip 1em plus
  0.5em minus 0.4em\relax Springer, 2017.

\end{thebibliography}
\end{document}